\documentclass[11pt]{amsart}
\newtheorem{Theorem}{Theorem}[section]
\newtheorem{Proposition}[Theorem]{Proposition}
\newtheorem{Lemma}[Theorem]{Lemma}

\newtheorem{Remark}[Theorem]{Remark}

\newtheorem{Assumption 2}[Theorem]{Assumption 2}

\numberwithin{equation}{section}

\usepackage{ulem}
\def\k#1{\kern#1em}
\def\Ib#1{{I\kern-.25em#1}}
\def\Ibb#1{{I\kern-.23em#1}}

\def\BB{\mathbb B}
\def\CC{{\mathbb C}}

\def\HH{{\mathbb H}}

\def\NN{{\mathbb N}}

\def\RR{{\mathbb{R}}}

\def\ZZ{{\mathbb Z}}
\def\vci{\vrule  width.02em height1.47ex depth-.0ex}
\def\11{{\rm\k{.2}\vci\k{-.37}1}}

\def\fin{{\begin{flushright}
\it{Q.E.D.}
\end{flushright}}}

\begin{document}

\address{Universit\'e de Bordeaux, Institut de Math\'ematiques, UMR CNRS 5251, F-33405 Talence Cedex}

\email{bachelot@math.u-bordeaux1.fr}

\title{The Dirac system on the Anti-de Sitter Universe}

\author{Alain BACHELOT}

\begin{abstract}
We investigate the global solutions of the Dirac equation on the
Anti-de-Sitter Universe. Since this space is not globally hyperbolic,
the Cauchy problem is not, {\it a priori}, well-posed. Nevertheless we
can prove that there exists unitary dynamics, but its uniqueness crucially depends
on the ratio beween the mass $M$ of the field and the cosmological
constant $\Lambda>0$ :
 it appears  a critical value,
  $\Lambda/12$, which plays a role similar to the Breitenlohner-Freedman
  bound for the scalar fields. When $M^2\geq \Lambda/12$ there exists
  a unique unitary
  dynamics. In opposite, for the light
  fermions satisfying $M^2<\Lambda/12$, we
  construct several asymptotic conditions at infinity, such that the
  problem becomes well-posed. In all the cases, the spectrum of the
  hamiltonian is discrete. We also prove a result of equipartition of
  the energy.
\end{abstract}
\maketitle
%%
%%
%\begin{center}
%{\bf Le syst\`{e}me de  Dirac sur l'univers Anti de Sitter}
%\end{center}
%\vspace{0.5cm}
%\begin{center}
%\begin{minipage}[r]{14cm}
%\footnotesize{\sc R\'esum\'e.}
%Nous cherchons des solutions globales de l'\'equation de Dirac dans l'univers Anti-de Sitter. Comme
%cet espace n'est pas globalement hyperbolique, le probl\`eme de Cauchy
%n'est pas, {\it a priori}, bien pos\'e. Nous montrons que c'est toutefois le cas quand la masse du champ est grande par rapport \`a la constante cosmologique. En revanche, pour les faibles masses, nous construisons diverses conditions asymptotiques \`a l'infini,
%rendant le probl\`eme bien pos\'e. Dans tous les cas, l'hamiltonien a
%un spectre discret. Nous \'etablisson \'egalement un r\'esultat
%d'\'equipartition de l'\'energie.
%\end{minipage}
%\end{center}

\pagestyle{myheadings}
\markboth{\centerline{\sc Alain Bachelot}}{\centerline{\sc The Dirac
    System on the Anti-de Sitter Universe}}

%%%%%%%%%%%%%%%%%%%%%%%%%%%%%%%%%%%%%%%%%%%%%%%%%%%%%%%%%%%%%%%%%%%%%%%%%%%%%%%%

%%%%%%%%%%%%%%  INTRODUCTION 
\section{Introduction}

There has been much recent interest for the field theory in the
covering space of the Anti-de-Sitter space-time $CAdS$, that appears as the ground state of gauged
supergravity group \cite{gibbons}. This lorentzian manifold is the
maximally symmetric solution of the Einstein equations with cosmological
constant $-\Lambda<0$ included. Its topology is $\RR_t\times\RR^3_X$,
but its causality is non trivial because it is non globally
hyperbolic : the Cauchy data on $\{t=0\}\times\RR^3$ determines the
evolution of the fields only in a region $D$, bounded by a null
hypersurface, called a Cauchy horizon. More precisely $D$ is defined by
$\mid t\mid
<\sqrt{\frac{3}{\Lambda}}\left(\frac{\pi}{2}-\arctan\left(\sqrt{\frac{\Lambda}{3}}\mid
    X\mid\right)\right)$. Thus we can think that to specify the physics apart from $D$,
we have to impose some asymptotic constraint at the infinity as $\mid
X\mid\rightarrow\infty$. Since the conformal boundary of
$CAdS$ is timelike, this condition can be considered as a
boundary condition. It is exactly the case for the massless fields
that are conformal invariant, and the massless scalar fields have been
studied in this spirit by Avis, Isham and Storey in \cite{avis}. For the massive fields the situation is
different because the gravitational potential relative to any origin
increases at large spatial distances from the origin. It causes
confinement of massive particles and prevents them from escaping to
infinity. In fact, the situation is rather subtle and depends on the
ratio between the mass of the field and the cosmological
constant. This phenomenon has been discovered by Breitenlohner and
Freedman  \cite{breit1}, \cite{breit2}, who have showed the existence
of  two critical values, the B-F bounds, for the scalar fields ; the first one
assures the positivity of the energy, and the second one assures the
uniqueness of the dynamics. In this paper, we establish a similar
result for the Dirac fields. The square of the mass of the spinors is compared with
a unique B-F bound that is equal to $\Lambda/12$. We shall see that
the physics of the heavy fermions ($M^2\geq\Lambda/12$)
is uniquely determined, but there exists a lot of possible dynamics for
the light fermions ($M^2<\Lambda/12$), involving the asymptotic forms,
at the $CAdS$ infinity, of classical
boundary conditions, local or non-local : {\it MIT-bag, Chiral,  APS} conditions, etc.
Of the mathematical point of view, the solutions of the initial value
problem are given in $D$ by the Leray-Hadamard theorem for the hyperbolic
equations $\partial_t\Psi=\mathbf H(X,\partial_X)\Psi$, and on
the whole space-time, we  solve the Cauchy problem by a spectral
approach, i.e. we look for the solutions formally given by
$\Psi(t)=e^{it\mathbf H}\Psi(0)$. Therefore we have to construct self-adjoint
extensions of the Dirac hamiltonian $\mathbf H(X,\partial_X)$. This
method was used by A. Ishibashi and R.M. Wald \cite{ishi1},
\cite{ishi2}, for the integer spin fields.\\

The paper is organized as follows. In part 2, we briefly describe the Anti-de-Sitter manifold, mainly the
different systems of  coordinates and the properties of the null and
time-like geodesics. The explicit forms of the Dirac equation on
$CAdS$ are described in section 3, and we state the main
result, Theorem \ref{maintheo}. We perform the spinoidal spherical
harmonics decomposition in the following part. The asymptotic
conditions and the self-adjoint extensions are discussed in the final
section. In a short appendix, we present un new proof of the B-F
bounds for the Klein-Gordon equation.\\

We end this introduction with some bibliographic information. 
Above all, we have to mention the
works treating of the scalar fields on $CAdS$,
\cite{avis}, \cite{breit1}, \cite{breit2}, \cite{wald}.
We
refer to \cite{gibbons}, \cite{haw}, \cite{oneill} for a
presentation of the Anti-de-Sitter universe. There are many
mathematical works on the one-half spin field on curved space-time,
in particular \cite{bachelot-motet}, \cite{haf}, \cite{ha-n},
\cite{melnyk}, \cite{JPN-Dirac}, \cite{JPNdiss}, \cite{JPN3/2}. The
gravitational potential plays the role of a variable mass that tends
to the infinity at the space infinity ; the rather similar Dirac
equation on Minkowski space with increasing potential has been
considered in \cite{kalf}, \cite{schmidt}, \cite{yamada}. The
litterature on the boundary value problems for the Dirac system is
huge ; among important contributions, we can cite \cite{bartnik}, \cite{booss},
\cite{bruning}, \cite{bunke}, \cite{grubb}, \cite{hijazi}. There are
few  papers concerning
the deep problem of the global existence of fields on the non-globally
hyperbolic lorentzian manifolds, in particular \cite{ba1},
\cite{choquet}, \cite{fr2}, \cite{ishi1}, \cite{ishi2}, \cite{segev}.

%%%%%%%%%%%%%%%%%%%%%%%%%%%%%%%%%%%%%%%%%%%%%%%%%%%%%%%%%%%%%%%%%%%%%%%%%%%%%%%%

\section{The Anti-de-Sitter space time}

Given $\Lambda>0$, the anti-de-Sitter space $AdS$ is defined as the quadric
$$
(X^1)^2+(X^2)^2+(X^3)^2-U^2-V^2=-\frac{3}{\Lambda}
$$
embedded in the flat 5-dimensional space $\RR^5$ with the metric
$$
ds^2=dU^2+dV^2-(dX^1)^2-(dX^2)^2-(dX^3)^2.
$$
$AdS$ is the maximally symmetric solution of Einstein's equations
with an attractive cosmological constant $-\Lambda<0$. To describe $AdS$
it is convenient to set
$$
U=R\cos\left(\sqrt{\frac{\Lambda}{3}}T\right),\;\;V=R\sin\left(\sqrt{\frac{\Lambda}{3}}T\right),
$$
then we can see that
$$
AdS=S^1_T\times \RR^3_{(X^1,X^2,X^3)},
$$
$$
ds^2_{AdS}=\frac{\Lambda}{3}R^2dT^2+dR^2-(dX^1)^2-(dX^2)^2-(dX^3)^2,\;\;R=\sqrt{(X^1)^2+(X^2)^2+(X^3)^2+\frac{3}{\Lambda}}.
$$
For constant $T$, the slice $\{T\}\times\RR^3$ is exactly the
3-dimensional hyperbolic space $\mathbb H^3$  that is the upper sheet
of the hyperboloid $(X^1)^2+(X^2)^2+(X^3)^2-W^2=-\frac{3}{\Lambda}$ in the Minkowski space
$\RR^4_{(X^1,X^2,X^3,W)}$ with the metric
$(dX^1)^2+(dX^2)^2+(dX^3)^2-dW^2$. It is useful to use the spherical
coordinates
$$
r=\sqrt{(X^1)^2+(X^2)^2+(X^3)^2}\in [0,\infty[,\;\; and\;\;if\;\;0<r,\;\;\omega=\frac{1}{r}(X^1,X^2,X^3)\in
S^2,
$$
for which the hyperbolic metric becomes
$$
ds^2_{\mathbb H^3}=\left(1+\frac{\Lambda}{3}r^2\right)^{-1}dr^2+r^2d\omega^2
$$
where $d\omega^2$ is the euclidean metric on the unit two-sphere $S^2$,
$$
d\omega^2=d\theta^2+\sin^2\theta d\varphi^2,\;\;0\leq\theta\leq\pi,\;\;0\leq\varphi<2\pi.
$$
We shall use the nice picture of the hyperbolic space, the so called
Poincar\'e ball. We introduce
$$
1\leq j\leq 3,\;\;x^j=\sqrt{\frac{\Lambda}{3}}\frac{1}{1+\sqrt{1+\frac{\Lambda}{3}r^2}}X^j,
$$
$$
\varrho=\sqrt{\frac{\Lambda}{3}}\frac{r}{1+\sqrt{1+\frac{\Lambda}{3}r^2}}\in
[0,1[,
$$
then $\mathbb H^3$ can be seen as the unit ball
$$
\mathbb B=\{ {\mathbf x}:=(x^1,x^2,x^2)\in\RR^3;\;\;\varrho^2=(x^1)^2+(x^2)^2+(x^3)^2< 1\}
$$
endowed with the metric
$$
ds^2_{\mathbb
  H^3}=\frac{3}{\Lambda}\frac{4}{\left(1-\varrho^2\right)}\left(d\varrho^2+\varrho^2d\omega^2\right),\;\;0\leq \varrho<1,\;\;\omega\in S^2.
$$

We note that the time coordinate $T$ is periodic, and this property
implies the existence of closed timelike curves. To avoid this
unpleasant fact, we replace $T\in S^1$ by $t\in\RR$, i.e. we change
the topology, and we consider in this paper the Universal Covering
Space of the anti-de-Sitter space-time, that is the lorentzian manifold 
$CAdS:=({\mathcal M}, g)$ defined by
$$
{\mathcal
  M}=\RR_t\times \RR^3_{(X^1,X^2,X^3)}=\RR_t\times \mathbb B_{(x^1,x^2,x^3)},
 $$
with the metric, 
\begin{equation*}
\begin{split}
g_{\mu\nu}dx^{\mu}dx^{\nu}&=\left(1+\frac{\Lambda}{3}r^2\right)dt^2-\left(1+\frac{\Lambda}{3}r^2\right)^{-1}dr^2-r^2d\omega^2,\;\;0\leq
r<\infty,\;\;\omega\in S^2,\\
&=\left(\frac{1+\varrho^2}{1-\varrho^2}\right)^2dt^2-\frac{3}{\Lambda}\frac{4}{\left(1-\varrho^2\right)}\left(d\varrho^2+\varrho^2d\omega^2\right),\;\;0\leq
\varrho<1,\;\;\omega\in S^2.
\end{split}
\end{equation*}
It will be useful to
introduce a third radial coordinate,
\begin{equation}
x=\arctan\left(\sqrt{\frac{\Lambda}{3}}r\right)=2\arctan\varrho.
  \label{xxx}
\end{equation}
Then the Anti-de-Sitter manifold can be described by :
$$
(t,x,\theta,\varphi)\in \RR\times[0,\frac{\pi}{2}[\times[0,\pi]\times[0,2\pi[,
 $$
$$
g_{\mu\nu}dx^{\mu}dx^{\nu}=\left(1+\tan^2x\right)\tilde{g}_{\mu\nu}dx^{\mu}dx^{\nu},
$$where $\tilde g$ is given by
$$
\tilde{g}_{\mu\nu}dx^{\mu}dx^{\nu}=dt^2-\frac{3}{\Lambda}\left(dx^2+\sin^2xd\theta^2+\sin^2x\sin^2\theta
    d\varphi^2\right).
 $$
Therefore, if the $3$-sphere $S^3$ is parametrized by
$(x,\theta,\varphi)\in [0,\pi]\times [0,\pi]\times [0,2\pi[$, and
$S^3_+$ is the upper hemisphere $ [0,\frac{\pi}{2}[_x\times
[0,\pi]_{\theta}\times [0,2\pi[_{\varphi}$, $CAdS$ can be considered as conformally equivalent to the
submanifold $\widetilde{\mathcal M}=\RR_t\times S^3_+$ of the Einstein cylinder $({\mathcal E},\tilde g)$,
\begin{equation}
{\mathcal E}:=\RR_t\times S^3,
  \label{einstein}
\end{equation}
and the crucial point is that the boundary $\partial\widetilde{\mathcal
  M}=\RR_t\times\left\{x=\frac{\pi}{2}\right\}\times
S^2_{\theta,\varphi}$ is time-like. Nevertheless, we should note that,
unlike the black-hole horizon of the Schwarzschild metric (that is a
characteristic submanifold of the Kruskal space-time), the time-like
infinity of $CAdS$, like the cosmological horizon of the De Sitter
universe, (or a  rainbow) is seen in the same way by any observer : since $CAdS$ is {\it frame-homogeneous}  (i.e. any Lorentz
frame on $CAdS$ can be carried to any other by the differential map of an
isometry of $CAdS$), no point is privileged.\\

Finally we recall that the null geodesics of $AdS$ are  straight lines
in $\RR^5_{(X^1,X^2,X^3,U,V)}$ and the timelike geodesics are
ellipses, intersection of $AdS$ with the 2-planes of $\RR^5$ passing
through the origin $0$. As consequence, $CAdS$ is geodesically complete, and time
oriented by the Killing vector field $\partial_t$, but its
causality is not at all trivial : (1)  given a point $P$ on the slice
$t=0$, the future-pointing null geodesics starting from $P$ form a
curving cone of which the boundary approches but does not reach the
slice $t=\frac{\pi}{2}\sqrt{\frac{3}{\Lambda}}$,  hence $CAdS$ is not
globally hyperbolic; (2) the
future-pointing timelike geodesics on $CAdS$ starting from $P$, all met a
conjugate point $Q$ at $t=\pi\sqrt{\frac{3}{\Lambda}}$, $P$ and  $Q$
project on antipodal points of $AdS$.  Therefore the time-like
geodesics on $CAdS$ can be parametrized by $(t,{\mathbf
  x}(t))_{t\in\RR}$, where the function $t\mapsto\mathbf{x}(t)$ is
$t$-periodic, with period $2\pi\sqrt{\frac{3}{\Lambda}}$. These unusual
properties yield important consequences for the propagation of the
fields : (1) suggests that we could have to add some condition at the
``infinity'' $S^2=\partial\mathbb B$ to solve an initial value
problem, at least for the massless  fermions ; nevertheless, since the
massive particles propagate along the time-like geodesics, (2) seems
to imply that such a condition is not necessary for the massive
fields. In fact, the situation is rather subtle and depends on the
ratio between the square of the mass of the fermion, and the
cosmological constant. We shall see that no
asymptotic constraint at the infinity is necessary for the heavy
spinors, but there are many possible physics for the light
masses. In all the cases, the spectrum
of the hamiltonian of the massive fields is discrete.

%%%%%%%%%%%%%%%%%%%%%%%%%%%%%%%%%%%%%%%%%%%%%%%%%%%%%%%%%%%%%%%%%%%%%%%%%%%%%%%%%%%%%%%%%%%
%%%%%%%%%%%%   DIRAC   %%%%%%%%%%%%
%%%%%%%%%%%%%%%%%%%%%%%%%%%%%%%%%%%%%%%%%%%%%%%%%%%%%%%%%%%%%%%%%%%%%%%%%%%%%%%%%%%%%%%%%%%
\section{The Dirac Equation on $CAdS$}
We consider the Dirac equation with mass $M\in\RR$ on a 3+1
dimensional lorentzian manifold $(\mathcal M,g)$ :
\begin{equation}
i\gamma^{\mu}_{(g)}\nabla_{\mu}\psi-M\psi=0.
\label{D}
\end{equation}

The notations are the following. $\nabla_{\mu}$ are the covariant
derivatives,
$\gamma^{\mu}_{(g)}$, $0\leq\mu\leq 3$, are the Dirac matrices, unique up to
a unitary transform, satisfying :
\begin{equation}
\gamma^{0*}_{(g)}=\gamma^0_{(g)},\;\;\gamma^{j*}_{(g)}=-\gamma^{j}_{(g)},\;1\leq j\leq
3,\;\;\gamma^{\mu}_{(g)}\gamma^{\nu}_{(g)}+\gamma^{\nu}_{(g)}\gamma^{\mu}_{(g)}=2g^{\mu\nu}{\mathbf
1}.
  \label{comanti}
\end{equation}
Here $A^*$ denotes the conjugate transpose of any complex matrix
$A$.
We  make the following choices for the Dirac matrices on the Minkowski
space time $\RR^{1+3}$ : $\gamma^{\mu}$ are the $4\times 4$ matrices of the Pauli-Dirac representation given 
for ${\mu}=0,1,2,3$ by :
 
\begin{equation*}
{\gamma^0}=\left(
\begin{array}{cc}
I&0\\
0&-I
\end{array}
\right), \;\;{\gamma^{j}}=\left(
\begin{array}{cc}
0&{\sigma}^{j}\\
-{\sigma}^{j}&0
\end{array}
\right),
\end{equation*}
where
\begin{equation*}
I=\left(
\begin{array}{cc}
1&0\\
0&1
\end{array}
\right),\;\;
{\sigma^1}=\left(
\begin{array}{cc}
1&0\\
0&-1
\end{array}
\right),\;\;
{\sigma^2}=\left(
\begin{array}{cc}
0&1\\
1&0
\end{array}
\right) 
,\;\;{\sigma^3}=\left(
\begin{array}{cc}
0&-i\\
i&0
\end{array}
\right).
\end{equation*}
We also introduce another Dirac matrix that plays an important role
in the boundary problems :
\begin{equation}
\gamma^5:=-i\gamma^0\gamma^1\gamma^2\gamma^3=\left(
\begin{array}{cc}
0&I\\
I&0
\end{array}
\right),
  \label{gamak}
\end{equation}
that satisfies
$$
\gamma^5\gamma^{\mu}+\gamma^{\mu}\gamma^5=0,\;\;0\leq\mu\leq 3.
 $$
We know that when the metric is spherically symmetric,
$$
g_{\mu\nu}dx^{\mu}dx^{\nu}=F(r)dt^2-\frac{1}{F(r)}dr^2-r^2\left(d\theta^2+\sin^2\theta
  d\varphi^2\right),
$$
then, if we choose the local orthonormal Lorentz frame $\{e_a^{\mu},a=0,1,2,3\}$  defined by
$$
e_{a}^{\;\;\mu}=\left\vert
  g^{\mu\mu}\right\vert^{\frac{1}{2}},\;\;if
\;\;\mu=a,\;\;e_{a}^{\;\;\mu}=0\;\;if\;\;\mu\neq a,
$$
the Dirac equation has the following form in $(t,r,\theta,\varphi)$ coordinates (see e.g. \cite{JPN-Dirac}, \cite{JPNdiss}, \cite{JPN3/2}):
\begin{equation*}
\Big\{iF^{-\frac{1}{2}}\gamma^0\frac{\partial}{\partial t}
+iF^{\frac{1}{2}}\gamma^1\left(\frac{\partial}{\partial r}+
\frac{1}{r}+\frac{F'}{4F}\right)+\frac{i}{r}\gamma^2\left(\frac{\partial}{\partial \theta}
+\frac{1}{2\tan \theta}\right)
+\frac{i}{r\sin \theta}\gamma^3\frac{\partial}{\partial \varphi}
-M\Big\}\psi=0.
\label{didi}
\end{equation*}
For the Anti-de-Sitter manifold we have
$$
F(r)=\left(1+\frac{\Lambda}{3}r^2\right),
$$
and it is convenient to make a first change of spinor ; we use the radial
coordinate (\ref{xxx}), and we put
\begin{equation}
\Phi(t,x,\theta,\varphi):=r\left(1+\frac{\Lambda}{3}r^2\right)^{\frac{1}{4}}\psi(t,r,\theta,\varphi).
\label{fi}
\end{equation}
Then we obtain  the Dirac equation on the
Anti-de-Sitter universe with the coordinates $t\in\RR$,
$x\in[0,\frac{\pi}{2}[$,  $\theta\in[0,\pi]$,  $\varphi\in[0,2\pi[$ :
\begin{equation}
\sqrt{\frac{3}{\Lambda}}\frac{\partial}{\partial t}\Phi
+\gamma^0\gamma^1\frac{\partial}{\partial x}\Phi+
\frac{1}{\sin x}\left[\gamma^0\gamma^2\left(\frac{\partial}{\partial \theta}
+\frac{1}{2\tan \theta} \right)
+\frac{1}{\sin \theta}\gamma^0\gamma^3\frac{\partial}{\partial \varphi}\right]\Phi
+\frac{i}{\cos x}M\sqrt{\frac{3}{\Lambda}}\gamma^0\Phi=0.
\label{EQu}
\end{equation}
Since the part of this differential operator involving $\partial_t$, $\partial_x$ is with
constant coefficients, the form of this equation is convenient to make a separation of
variables by using the generalized spin spherical harmonics. But this
decomposition has an inconvenience : since
the one-half spin harmonics are not smooth functions on $S^2$, the
functional framework involves spaces that are different from the usual
Sobolev spaces on $S^2$ as we shall see in the following
part. It will also be useful to write the Dirac equation with the
 coordinates $(t,\varrho,\theta,\varphi)\in\RR\times[0,1[\times[0,\pi]\times[0,2\pi[$. We put
\begin{equation*}
\uwave\Phi(t,\varrho,\theta,\varphi):=\Phi(t,x,\theta,\varphi),
\end{equation*}
and the Dirac equation becomes :
\begin{equation*}
\sqrt{\frac{3}{\Lambda}}\frac{\partial}{\partial t}\uwave\Phi
+\left(\frac{1+\varrho^2}{2}\right)\gamma^0\left[\gamma^1\frac{\partial}{\partial \varrho}+
\frac{1}{\varrho}\gamma^2\left(\frac{\partial}{\partial \theta}
+\frac{1}{2\tan \theta} \right)
+\frac{1}{\varrho\sin \theta}\gamma^3\frac{\partial}{\partial \varphi}
+\frac{2iM}{1-\varrho^2}\sqrt{\frac{3}{\Lambda}}\right]\uwave\Phi=0.
\end{equation*}
In the part involving $\gamma^1$, $\gamma^2$, $\gamma^3$, we recognize
the usual Dirac operator in spherical coordinates on $\RR^3$ with the
euclidean metric. This is a nice way  to get the Dirac operator on
$CAdS$ in cartesian coordinates. Adapting an approach of
\cite{shishkin}, we introduce 
$$
a:=\frac{1}{2}\left(I-\gamma^1\gamma^2-\gamma^2\gamma^3-\gamma^3\gamma^1\right),
$$
\begin{equation}
S(\theta,\varphi):=e^{\frac{\varphi}{2}\gamma^1\gamma^2}e^{\frac{\theta}{2}\gamma^3\gamma^1}a.
  \label{SSS}
\end{equation}
We easily check that
$$
aa^*=I,\;\;SS^*=I,
$$
$$
\gamma^1a=a\gamma^2,\;\;\gamma^2a=a\gamma^3,\;\;\gamma^3a=a\gamma^1.
 $$
We put
$$
\uwave\gamma^1(\varrho,\theta):=\gamma^1,\;\;\uwave\gamma^2(\varrho,\theta):=\frac{1}{\varrho}\gamma^2,\;\;\uwave\gamma^3(\varrho,\theta):=\frac{1}{\varrho\sin\theta}\gamma^3,
 $$
\begin{equation*}
\left\{
\begin{split}
\tilde\gamma^1(\varrho,\theta,\varphi):=&\cos\varphi\sin\theta\gamma^1+\sin\varphi\sin\theta\gamma^2+\cos\theta\gamma^3,\\
\tilde\gamma^2(\varrho,\theta,\varphi):=&\frac{1}{\varrho}\left(\cos\varphi\cos\theta\gamma^1+\sin\varphi\cos\theta\gamma^2-\sin\theta\gamma^3\right),\\
\tilde\gamma^3(\varrho,\theta,\varphi):=&\frac{1}{\varrho\sin\theta}\left(-\sin\varphi\gamma^1+\cos\varphi\gamma^2\right).
\end{split}
\right.
\end{equation*}
Tedious calculations give :
$$
1\leq j\leq 3,\;\;S(\theta,\varphi)\uwave\gamma^j(\varrho,\theta)=\tilde\gamma^j(\varrho,\theta,\varphi)S(\theta,\varphi).
$$
The cartesian coordinates ${\mathbf x}:=(x^1,x^2,x^3)$ on $\mathbb B$ being
\begin{equation}
x^1=\varrho\cos\varphi\sin\theta,\;\;x^2=\varrho\sin\varphi\sin\theta,\;\;x^3=\varrho\cos\theta,
  \label{cart}
\end{equation}
we define the spinors $\Psi$, $\tilde\Phi$ on $\mathbb B$ by the relations
$$
\Psi(x^1,x^2,x^3)=\tilde\Phi(\varrho,\theta,\varphi):=\frac{1}{\varrho}S(\theta,\varphi)\uwave\Phi(\varrho,\theta,\varphi),
$$
and the Dirac operators 
\begin{equation*}
\left\{
\begin{split}
{\mathbb D}:&=\gamma^1\frac{\partial}{\partial
  x^1}+\gamma^2\frac{\partial}{\partial
  x^2}+\gamma^3\frac{\partial}{\partial x^3},\\
\tilde{\mathbb D}:&=\tilde\gamma^1\frac{\partial}{\partial
  \varrho}+\tilde\gamma^2\frac{\partial}{\partial
  \theta}+\tilde\gamma^3\frac{\partial}{\partial \varphi},\\
\uwave{\mathbb D}:&=\uwave\gamma^1\frac{\partial}{\partial
  \varrho}+\uwave\gamma^2\left(\frac{\partial}{\partial
  \theta}+\frac{1}{2\tan\theta}\right)+\uwave\gamma^3\frac{\partial}{\partial \varphi}.
\end{split}
\right.
\end{equation*}
We omit the  direct calculus that gives the links between these operators :

\begin{Lemma}
 $$
\left({\mathbb D}\Psi\right)(x^1,x^2,x^3)=\left(\tilde{\mathbb D}\tilde\Phi\right)(\varrho,\theta,\varphi)=\frac{1}{\varrho}S(\theta,\varphi)\left(\uwave{\mathbb D}\uwave\Phi\right)(\varrho,\theta,\varphi).
   $$
\end{Lemma}

We denote $\mathbf S$ the operator that relates the spinors in
cartesian and spherical coordinates :
\begin{equation}
\mathbf{S}: \Phi\mapsto\mathbf{S}\Phi=\Psi,\;\;\Psi(x^1,x^2,x^3):
=\frac{1}{\tan\left(\frac{x}{2}\right)}S(\theta,\varphi)\Phi(x,\theta,\varphi).
  \label{S!}
\end{equation}
Then, if $\Phi(t,.)$ is a solution of (\ref{EQu}),  the Dirac equation
satisfied by $\Psi(t,.):={\mathbf S}\Phi(t,.)$ for $t\in\RR$, ${\mathbf
  x}=(x^1,x^2,x^3)\in\mathbb B$ has the form :
\begin{equation}
\sqrt{\frac{3}{\Lambda}}\gamma^0\frac{\partial}{\partial
  t}\Psi+\left(\frac{1+\varrho^2}{2}\right)\left[\gamma^1\frac{\partial}{\partial x^1}+\gamma^2\frac{\partial}{\partial x^2}+\gamma^3\frac{\partial}{\partial x^3}+\frac{2iM}{1-\varrho^2}\sqrt{\frac{3}{\Lambda}}\right]\Psi=0.
  \label{eqcar}
\end{equation}

Since the charge of the spinor is the formally conserved $L^2$ norm,
it is natural to introduce the Hilbert space :
\begin{equation}
{\mathbf
  L}^2:=\left[L^2\left({\mathbb B},\frac{2}{1+\varrho^2}{\mathbf{dx}}\right)\right]^4,
\label{espldd}
\end{equation}
and given $\Psi_0\in{\mathbf L}^2$ we want to solve the initial
problem, {\it i.e.}  to find a unique
\begin{equation}
\Psi\in C^0(\RR_t;{\mathbf L}^2)
  \label{reg}
\end{equation}
 solution of (\ref{eqcar})
 satisfying :
\begin{equation}
\Psi(t=0,.)=\Psi_0(.),
\label{CI}
\end{equation}
and  the conservation law :
\begin{equation}
\forall t\in\RR,\;\;\parallel\Psi(t)\parallel_{\mathbf
  L^2}=\parallel\Psi_0\parallel_{\mathbf L^2}.
  \label{conserve}
\end{equation}
Moreover, since
$\frac{\partial}{\partial t}$ is a Killing vector field on $CAdS$, it is
natural to assume that 
\begin{equation}
t\in\RR\longmapsto \left(\Psi_0\mapsto \Psi(t)\right),
  \label{group}
\end{equation}
is a group acting on ${\mathbf L}^2$.
Therefore we look for
strongly continuous unitary groups $U(t)$ on ${\mathbf L}^2$ that
solve (\ref{EQu}). According to the Stone theorem, the problem
consists in finding self-adjoint realizations on $\mathbf L^2$ of the differential
operator
\begin{equation}
{\mathbf H}_M:=i\left(\frac{1+\varrho^2}{2}\right)\gamma^0\left[\gamma^1\frac{\partial}{\partial x^1}+\gamma^2\frac{\partial}{\partial x^2}+\gamma^3\frac{\partial}{\partial x^3}+\frac{2iM}{1-\varrho^2}\sqrt{\frac{3}{\Lambda}}\right],
  \label{dirop}
\end{equation}
with domain
\begin{equation}
D({\mathbf H}_M)=\left\{\Psi\in{\mathbf L}^2;\;{\mathbf H}_M\Psi\in{\mathbf L}^2\right\},
  \label{domdi}
\end{equation}
by adding suitable constraints at the {\it CAdS} infinity $\varrho=1$. The answer crucially depends on the mass of the
spinor.\\

First we discuss the massless case. When $M=0$, the Dirac system
is conformal invariant, and it is equivalent to solving the Cauchy
problem in the half of the Einstein cylinder, $\RR_t\times
S^3_+$. Therefore we can extend the initial data from the hemisphere
$S^3_+$ to the whole sphere $S^3$, and  solve the Cauchy problem on the
Einstein cylinder $\RR_t\times
S^3$.
This is tantamount to solving the equation (\ref{eqcar}) on
$\RR_t\times\RR^3_{\mathbf x}$, instead of $\RR_t\times\mathbb{B}_{\mathbf x}$.
 This approach
was used by
S.J. Avis, C.J. Isham, D. Storey \cite{avis} for the scalar field, and
later, by Y. Choquet-Bruhat for the Yang-Mills-Higgs equations
\cite{choquet}.
 By this way, we impose no boundary condition at the $CAdS$
infinity, or, in other words, a ``perfectly transparent'' boundary
condition, and  we easily obtain global solutions on $CAdS$. We
have to remark that since there
exists a lot of ways to extend the initial data, such a solution is
not uniquelily determined by the Cauchy data on $S^3_+$. Moreover the
effect of this ``perfectly transparent'' condition is to recirculate
the energy : the
conserved charge is the $L^2$-norm on $S^3$ while the $L^2$-norm on
$S^3_+$ in changing in time, and so (\ref{conserve}) is not satisfied.
 In order to assure the conservation  (\ref{conserve}), we can take
 another route, and impose some ``reflecting'' boundary conditions on
 $\{x=\frac{\pi}{2}\}\times S^2$. In \cite{avis}, several conditions are
 discussed for the scalar massless field. 
For the Dirac equation, we note that when $M=0$, equation (\ref{eqcar}) has  smooth
coefficients up to the boundary $\mid\mathbf x\mid=1$. Therefore, in the massless case, we deal with a classical mixed hyperbolic problem, and different
boundary conditions for the Dirac system with regular potential are
well known (see e.g. \cite{bartnik}, \cite{booss}, \cite{bruning},
\cite{bunke}, \cite{grubb}, \cite{hijazi}).
We recall an important local boundary condition  for the Dirac
spinors defined on some open domain $\Omega$ of the space-time, the so
called generalized {\it  MIT-bag} condition :
\begin{equation*}
n_{\mu}\gamma^{\mu}\Psi(t,x^1,x^2,x^3)=ie^{i\alpha\gamma^5}\Psi(t,x^1,x^2,x^3),\;\;(t,x^1,x^2,x^3)\in\partial\Omega,
\end{equation*}
where $n^{\mu}$ is the outgoing normal quadrivector at
$\partial\Omega$ and $\alpha\in\RR$ is the chiral angle. When
$\alpha=0$ this is the {\it MIT-bag} condition for the hadrons  and
when $\alpha=\pi$ this is the {\it Chiral} condition. Another fundamental
boundary condition is the non-local {\it APS} condition introduced by  M.F. Atiyah, V. K. Patodi, and
I. M. Singer (see e.g. \cite{booss}) and defined by
\begin{equation*}
{\mathbf 1}_{]0,\infty[}\left(D_{\partial\Omega}\right)\Psi=0\;\;on\;\;\partial\Omega,
\end{equation*}
where $ D_{\partial\Omega}$ is the Dirac operator on
$\partial\Omega$.
 More recently, O. Hijazi, S. Montiel,
A. Roldan \cite{hijazi} have introduced the
{\it mAPS} condition :
\begin{equation*}
{\mathbf 1}_{]0,\infty[}\left(D_{\partial\Omega}\right)\left(Id-n_{\mu}\gamma^{\mu}\right)\Psi=0\;\;on\;\;\partial\Omega.
\end{equation*}

For
$\Omega=\RR_t\times\mathbb B$, these boundary conditions become
\begin{equation}
{\mathcal B}\Psi(t,\omega)=0,\;\;(t,\omega)\in\RR\times S^2,
  \label{condil}
\end{equation}
where 
\begin{equation}
MIT-bag\;condition:\;\;{\mathcal B}_{MIT}=\tilde\gamma^1+iId,
  \label{mitcar}
\end{equation}
\begin{equation}
Chiral\;condition:\;\;{\mathcal B}_{CHI}=\tilde\gamma^1-iId,
  \label{chicar}
\end{equation}
\begin{equation}
APS\;condition:\;\;{\mathcal B}_{APS}={\mathbf
  1}_{]0,\infty[}\left(D_{S^2}\right),
  \label{apspsi}
\end{equation}
\begin{equation}
mAPS\;condition:\;\;{\mathcal B}_{mAPS}={\mathbf
  1}_{]0,\infty[}\left(D_{S^2}\right)\left(\tilde\gamma^1+Id\right),
  \label{mapspsi}
\end{equation}
where $D_{S^2}$ is the intrinsic Dirac operator on the two-sphere :
$$
\widetilde{D_{S^2}\Psi}=i\gamma^0\left(\tilde\gamma^2\frac{\partial}{\partial
  \theta}+\tilde\gamma^3\frac{\partial}{\partial \varphi}\right)\tilde\Phi.
 $$
We conclude that there exists many unitary dynamics for the massless
spin-$\frac{1}{2}$ field on $CAdS$, that we can easily construct by
solving (\ref{eqcar}) with $M=0$, (\ref{CI}), (\ref{condil}), by
invoking the classical theorems on
the mixed hyperbolic problems. In consequence, our work is mainly
concerned with the massive field, and in the sequel, we  consider only
this case.\\

When $M\neq 0$ the situation is very different because the potential
blows up as $\varrho\rightarrow 1$. The analogous situation of the
infinite mass at the infinity of the Minkowski space has been
investigated in \cite{kalf}, \cite{schmidt}.  In our case, the key
result is the asymptotic behaviour, near the boundary, of the spinors
of $D(\mathbf H_M)$. We note that it is sufficient to consider only the case of the positive mass,
because the chiral transform
$$
\Psi\longrightarrow \gamma^5\Psi
$$
changes $M$ into $-M$ since we have
$$
\gamma^5\mathbf{H}_M\gamma^5=\mathbf{H}_{-M}.
$$
We remark that the {\it MIT-bag} and the {\it Chiral} conditions are
exchanged by the chiral transform, and the {\it APS} condition is chiral
invariant.

%%%%%%%%%%%%%%%%%%%%%%%%%%%%%%%%%%%%%%%%%%%%%%%%%%%%%%%%%%%%%%%%%%%%%%%%%%%%%%

%%%%%%%%%%%%%%%%%%%%%%%%%%%%%  THEO  BOUNDARY  %%%%%%%%%%%%%%%%%%%%

\begin{Theorem} Let $\Psi$ be in $D\left(\mathbf H_M\right)$with
  $M\in\RR^*$. Then
\begin{equation}
\Psi\in
\left[C^0\left(]0,1[_{\varrho};H^{\frac{1}{2}}(S^2_{\omega}\right)\right]^4,
  \label{regfip}
\end{equation}
\begin{equation}
\int_0^1\parallel
\Psi(\varrho\omega)\parallel_{H^1(S^2_{\omega})}^2\varrho d\varrho\leq
\parallel \mathbf H_M\Psi\parallel_{\mathbf L^2}^2.
  \label{estcar}
\end{equation}
When $M^2>\frac{\Lambda}{12}$, we have
\begin{equation}
\parallel \Psi(\varrho\omega)\parallel_{L^2(S^2_{\omega})}=O\left(\sqrt{1-\varrho}\right),\;\;\varrho\rightarrow 1.
  \label{dekop}
\end{equation}
When
$
M^2=\frac{\Lambda}{12}$,
we have 
\begin{equation}
\parallel
\Psi(\varrho\omega)\parallel_{L^2(S^2_{\omega})}=O\left(\sqrt{\left(\varrho-1\right)\ln\left(1-\varrho\right)}\right),\;\;\varrho\rightarrow
1.
  \label{dekup}
\end{equation}
When $0<M^2<\frac{\Lambda}{12}$,
we put
$
m:=M\sqrt{\frac{3}{\Lambda}}$,
and there exists $\Psi_-\in \left[H^{\frac{1}{2}}(S^2)\right]^4$,
 $\Psi_{+}\in \left[L^2(S^2)\right]^4$, and $\psi\in
\left[C^0\left([0,1]_{\varrho};L^2(S^2_{\omega})\right)\right]^4$
satisfying
\begin{equation}
\Psi(\varrho\omega)=
\left(1-\varrho\right)^{-m}
\Psi_-(\omega)
+
\left(1-\varrho\right)^{m}
\Psi_+(\omega)+\psi(\varrho\omega),
\label{dqfp}
\end{equation}
\begin{equation}
\tilde\gamma^1\Psi_-+i\Psi_-=0,\;\;\tilde\gamma^1\Psi_+-i\Psi_+=0,
  \label{polar}
\end{equation}
\begin{equation}
\parallel \psi(\varrho\omega)\parallel_{L^2(S^2_{\omega})}=o\left(\sqrt{1-\varrho}\right),\;\;\varrho\rightarrow 1.
  \label{faqp}
\end{equation}
Conversely, for any  $\Psi_-\in \left[H^{\frac{1}{2}+m}(S^2)\right]^4$,
$\Psi_{+}\in \left[H^{\frac{1}{2}-m}(S^2)\right]^4$ satisfying  (\ref{polar}),
 there exists $\Psi \in D(\mathbf H_M)$
satisfying (\ref{dqfp}) and (\ref{faqp}).
  \label{behcar}
\end{Theorem}

\begin{Remark}
We shall see that (\ref{dekop}) can be improved and when
$M^2>\frac{\Lambda}{12}$ the elliptic estimate below (\ref{ellip})
implies that $\int_0^1 \parallel \Psi(\varrho\omega)\parallel^2_{L^2(S^2_{\omega})}\frac{d\varrho}{(1-\varrho)^2}<\infty$. When $M^2\geq\frac{\Lambda}{12}$, then $\Psi\in
\left[C^0(]0,1]_{\varrho}; L^2(S^2_{\omega})\right]^4$, but the trace
of $\Psi$ on $\partial\mathbb B$ does not exist for
$M^2<\frac{\Lambda}{12}$. Moreover we see with (\ref{estcar}) that
when $M\neq 0$, $\mathbf H_M\Psi=0$
implies $\Psi=0$.The situation is different when $M=0$ : we have  $\Psi\in
\left[C^0\left(]0,1]_{\varrho};H^{-\frac{1}{2}}(S^2_{\omega}\right)\right]^4$
for $\Psi\in D({\mathbf H}_0)$,
and this result is optimal : there exists $\Psi\in {\mathbf L}^2$,
$\Psi\neq 0$,  with
${\mathbf H}_0\Psi=0$ and $\Psi(\omega)\in
\left[H^{-\frac{1}{2}}(S^2_{\omega})\right]^4\setminus
\cup_{s>-\frac{1}{2}}\left[H^{s}(S^2_{\omega})\right]^4$.
\label{rembp}
\end{Remark}

We note that when $M^2\geq \frac{\Lambda}{12}$, the elements of
the domain of $\mathbf H_M$ satisfy the homogeneous Dirichlet Condition on
$\partial\mathbb B$. We shall see that $\mathbf H_M$ is self-adjoint. In opposite, when
$0<M<\sqrt{\frac{\Lambda}{12}}$, the trace of $\Psi$ on
$\partial\mathbb B$ is not defined, the leading term
$(1-\varrho)^{-m}\Psi_-$ satisfies the {\it MIT-bag} Condition and  the next
term $(1-\varrho)^m\Psi_+$ satisfies the {\it Chiral} Condition (and
the converse for $-\sqrt{\frac{\Lambda}{12}}<M<0$). We
introduce natural generalizations of the classic boundary conditions
in terms of asymptotic behaviours near $S^2$ :
\begin{equation}
\parallel{\mathcal B}\Psi(\varrho\omega)\parallel_{L^2(S^2_{\omega})}=o\left(\sqrt{1-\varrho}\right),
  \label{assco}
\end{equation}
and we consider the operators $\mathbb H_{\mathcal B}$,
$\mathcal B=\mathcal B_{MIT},\;\mathcal B_{CHI},\;\mathcal B_{APS},\;\mathcal B_{mAPS}$,  defined
as the differential operator $\mathbf H_M$ endowed with the domain
$$
D\left(\mathbb H_{\mathcal B}\right):=\left\{\Psi\in D(\mathbf H_M);\;\parallel{\mathcal B}\Psi(\varrho\omega)\parallel_{L^2(S^2_{\omega})}=o\left(\sqrt{1-\varrho}\right)\right\}.
 $$
We remark that (\ref{dqfp}), (\ref{polar}) and (\ref{faqp}) imply :
$$
D\left(\mathbb H_{\mathcal B_{MIT}}\right):=\left\{\Psi\in D(\mathbf
  H_M);\;\Psi_+=0\; if \; M>0,\;\Psi_-=0\; if \; M<0\right\},
 $$
 $$
D\left(\mathbb H_{\mathcal B_{CHI}}\right):=\left\{\Psi\in D(\mathbf
  H_M);\;\Psi_-=0\; if\; M>0,\;\Psi_+=0\; if\; M<0\right\}.
$$
$$
D\left(\mathbb H_{\mathcal B_{APS}}\right)=D\left(\mathbb H_{\mathcal
    B_{mAPS}}\right)
=
\left\{\Psi\in D(\mathbf
  H_M);\;{\mathbf
  1}_{]0,\infty[}\left(D_{S^2}\right)\Psi_+={\mathbf
  1}_{]0,\infty[}\left(D_{S^2}\right)\Psi_-=0\right\}.
$$

We now construct a large family of asymptotic conditions, generalizing
the previous one, by imposing a linear relation between $\Psi_-$ and $\Psi_+$. If we denote
$\Psi_{\pm}=^t(\psi_{\pm}^1,\psi_{\pm}^2,\psi_{\pm}^3,\psi_{\pm}^4)$,
the constraints of polarization (\ref{polar}) allow to express
$\psi^{3,4}_{\pm}$ by using $\psi^{1,2}_{\pm}$ :
\begin{equation*}
\left(
\begin{array}{c}
\psi_{\pm}^3(\omega)\\
\psi_{\pm}^4(\omega)
\end{array}
\right)
=
\pm i
\pmb{\omega.\sigma}
\left(
\begin{array}{c}
\psi_{\pm}^1(\omega)\\
\psi_{\pm}^2(\omega)
\end{array}
\right),\;\;
\pmb{\omega.\sigma}:=
\sum_1^3\omega^j\sigma^j.
\end{equation*}
We consider two densely defined self-adjoint operators $(\mathbf
A^{\pm}, D(\mathbf A^{\pm}))$ on $L^2(S^2)\times L^2(S^2)$, satisfying
\begin{equation}
D(\mathbf A^+)=L^2(S^2)\times L^2(S^2),\;\;D(\mathbf A^-)\supset
H^{\frac{1}{2}}(S^2)\times H^{\frac{1}{2}}(S^2),
  \label{dapm}
\end{equation}
\begin{equation}
\mathbf A^{\pm}\left(C^{\infty}(S^2)\times
  C^{\infty}(S^2)\right)\subset
H^{\frac{1}{2}\pm m}(S^2)\times H^{\frac{1}{2}\pm m}(S^2).
  \label{capm}
\end{equation}
As an example, we can choose $\mathbf A^-$ any hermitian matrix of
$H^{\frac{1}{2}}(S^2;\CC^{2\times 2})$, and $\mathbf A^+$ any hermitian matrix of
$H^{\frac{1}{2}+m}\cap L^{\infty}(S^2;\CC^{2\times 2})$.
We define the operators $(\mathbb H_{\mathbf A^{+}}, D(\mathbb
H_{\mathbf A^{+}}))$, $(\mathbb H_{\mathbf A^{-}}, D(\mathbb
H_{\mathbf A^{-}}))$, where
\begin{equation*}
D\left(\mathbb
H_{\mathbf A^{\pm}}\right):=\left\{\Psi\in D(\mathbf
  H_M);\;\left(
\begin{array}{c}
\psi_{\mp}^1\\
\psi_{\mp}^2
\end{array}
\right)=\mathbf A^{\pm}\left(
\begin{array}{c}
\psi_{\pm}^1\\
\psi_{\pm}^2
\end{array}
\right)\right\}.
\end{equation*}
For $\mathbf A^-=\mathbf A^+=0$, we obviously have $\mathbb H_{\mathbf A^{\mp}}=\mathbb H_{\mathcal
  B_{MIT}}$, $\mathbb H_{\mathbf A^{\pm}}=\mathbb H_{\mathcal
  B_{CHI}}$ if $\pm M>0$. Furthermore, the chiral transform $\Psi\rightarrow\gamma^5\Psi$
leads to the exchanges  $M\rightarrow-M$, $\mathbb H_{\mathcal
  B_{MIT}}\rightarrow\mathbb H_{\mathcal B_{CHI}}$,
 $\mathbb H_{\mathcal
  B_{CHI}}\rightarrow\mathbb H_{\mathcal B_{MIT}}$,
$\mathbb H_{\mathcal
  B_{APS}}\rightarrow\mathbb H_{\mathcal B_{APS}}$,
$\mathbb H_{\mathcal
  B_{\mathbf A^{\pm}}}\rightarrow\mathbb H_{\mathcal{B}_{\pmb{\omega.\sigma}{\mathbf A^{\pm}}\pmb{\omega.\sigma}}}$.
\\

We now state the main theorem of this paper.
%%%%%%%%%%%%%%%%%%%%%%%%%%%%%%%%%%
%%%%%%%%%%%%%%   THEO   %%%%%%%%%%%
%%%%%%%%%%%%%%%%%%%%%%%%%%%%%%%%%%%%%

\begin{Theorem}[\makebox{\bf Main result}]
Given $M\in\RR^*$, we consider the massive  Dirac hamiltonian $
\mathbf H_M$  defined by (\ref{dirop}), (\ref{domdi}).
When
$
M^2\geq \frac{\Lambda}{12}$,
$\mathbf H_M$ is essentially self-adjoint on
$\left[C^{\infty}_0(\mathbb B)\right]^4$, and if
$
M^2> \frac{\Lambda}{12}$,
then
$
D\left(\mathbf H_M\right)=\left[H^1_0(\mathbb B)\right]^4,
$
and for all $\Psi\in D\left(\mathbf H_M\right)$, we have the following  elliptic
estimate :
\begin{equation}
\sqrt{\frac{\Lambda}{12}}
\parallel \mathbf H_M\Psi \parallel_{\mathbf L^2}\geq
\left(\mid M\mid-\sqrt{\frac{\Lambda}{12}}\right)\parallel\nabla_{\mathbf x}\Psi\parallel_{\mathbf L^2}.
  \label{ellip}
\end{equation}

When
$
M^2<\frac{\Lambda}{12}$,
$\mathbb H_{\mathbf A^+}$, $\mathbb H_{\mathbf A^-}$,
$\mathbb H_{\mathcal B_{APS}}$, $\mathbb H_{\mathcal B_{mAPS}}$ are
self-adjoint on $\mathbf L^2$, and $\mathbb H_{\mathcal
  B_{APS}}=\mathbb H_{\mathcal B_{mAPS}}$.

The resolvent of any self-adjoint realization of $\mathbf H_M$, $M\in\RR^*$,
is compact on $\mathbf L^2$, and so, the spectrum of these operators is discrete.

  \label{maintheo}
\end{Theorem}

We see that $\frac{\Lambda}{12}$ is an important critical value. It plays
exactly the same role that the bounds that  Breitenlohner and Freedman have
discovered for the scalar massive fields \cite{breit1}, \cite{breit2}. We recall
that these authors have considered the Klein-Gordon equation 
$
\mid g\mid^{-\frac{1}{2}}\partial_{\mu}\left(\mid g\mid^{\frac{1}{2}}g^{\mu\nu}\partial_{\nu}u\right)-\alpha\frac{\Lambda}{3}u=0,
$
for which $\alpha=2$ corresponds to the massless case. By a sharp
analysis of the modes, they have established, among other results, that: (i) the natural energy is positive when $\alpha\leq
9/4$ , in particular
for the light tachyons associated with $2<\alpha<9/4$; (ii) the dynamics is unique when $\alpha \leq 5/4$ ; (iii) there
exists a lot of unitary dynamics when $5/4<\alpha<9/4$. For a sake of
completeness we give in the one-page Appendix, a new and very simple proof of
these results, based on a Hardy estimate and on the Kato-Rellich
theorem. For the spin-$\frac{1}{2}$ field with real mass, the most important conserved
quantity is the $L^2$-norm that is always positive, hence one bound
will suffice to distinguish the different cases : it is
$\frac{\Lambda}{12}$. We have to emphasize that this value was already
presented in the discussion of the massive $OSp(1,4)$ scalar multiplet
in \cite{breit1}, \cite{breit2}. This multiplet consists of a Dirac
spinor with mass $M$, and two Klein-Gordon fields for which $\alpha=
2\pm M\sqrt{3/\Lambda}-3M^2/\Lambda$. We can easily check that $\alpha\leq
9/4$ for any $M\in\RR$, and $\alpha\leq 5/4$ iff $M^2\geq
\Lambda/12$. Therefore our own result is coherent with this particular
model of de Sitter supersymmetry : the constraints for the uniqueness
of the dynamics are simultaneously satisfied for the spin field and
the scalar fields. The case $\alpha>9/4$ describes the heavy tachyons in $CAdS$,
and corresponds to the case of an {\it imaginary} mass for the Dirac
field. This regime seems to be unphysical since the energy of a scalar
tachyon is not positive, and the $L^2$-norm of a spin-$\frac{1}{2}$
field with an imaginary mass is not conserved. Of the mathematical
point of view, it is doubtful that the global Cauchy problem with
these parameters is well posed,
and of the physical point of view, we could suspect that the $AdS$
background is not stable with respect to the fluctuations of such
fields. We do not adress this situation in this paper.
\\

 We now turn over to  the Cauchy problem.

\begin{Theorem}

Given $\Psi_0\in{\mathbf L}^2$, there exist solutions of (\ref{eqcar}),
(\ref{reg}), (\ref{CI}), and all the solutions are equal for
\begin{equation}
(t,\mathbf x)\in\RR\times\mathbb B,\;\;\mid t\mid <\sqrt{\frac{3}{\Lambda}}\left(\frac{\pi}{2}-2\arctan\varrho\right).
\end{equation}
When
$
M^2\geq \frac{\Lambda}{12}$,
the Cauchy problem (\ref{eqcar}),
(\ref{reg}), (\ref{CI}) has a unique solution. This solution satisfies (\ref{conserve}).

  \label{theoevol}
\end{Theorem}

We achieve this part  with a result of equipartition of the energy. We
know, \cite{equipart}, that the solutions $\Psi\in C^0(\RR_t; L^2(\RR^3;\CC^4))$ of the massive Dirac equation on the
Minkowski space-time, satisfy
$$
\lim_{\mid t\mid\rightarrow\infty}\int_{\RR^3}\Psi^*\gamma^0\gamma^5\Psi(t,\mathbf x)d\mathbf x=0.
$$
Since the spectrum of the possible hamiltonians for the massive
fermions on $CAdS$ is discrete, we cannot expect such an asymptotic
behaviour. Nevertheless, we establish the existence of a similar
limit, in the weaker sense of Cesaro : 

\begin{Theorem}
Let $\Psi\in C^0(\RR_t;\mathbf L^2)$ be a solution of (\ref{eqcar}),
given by $\Psi(t)=e^{it\sqrt{\frac{\Lambda}{3}}\mathbb H}\Psi(0)$ where
$\mathbb H$ is a self-adjoint realization of $\mathbf H_M$,
$M\in\RR^* $. Then we have :
\begin{equation}
\lim_{T\rightarrow\infty}\frac{1}{T}\int_0^T\int_{\mathbb
  B}\Psi^*\gamma^0\gamma^5\Psi(t,\mathbf x)d\mathbf x dt=0.
  \label{equipa}
\end{equation}
  \label{equipat}
\end{Theorem}

%%%%%%%%%%%%%%%%%%%%%%%%%%%%%%%%%%%%%%%%%%%%%%%%%%%%%%%%%%%%%%%%%
The proofs of these results  are presented in the parts V and VI. They
are made much easier
by the use of the spherical coordinates. Operator $\mathbf S$, defined
by (\ref{S!}), that relates the spinors within the two systems of
coordinates,  is an isometry from 
\begin{equation}
{\mathcal
  L}^2:=\left[L^2\left([0,\frac{\pi}{2}[_x\times[0,\pi]_{\theta}\times[0,2\pi[_{\varphi},\sin\theta dx d\theta d\varphi\right)\right]^4,
\label{esplde}
\end{equation}
onto $\mathbf L^2$, and satisfies the intertwining relation
\begin{equation}
{\mathbf H}_M{\mathbf S}={\mathbf S}H_m
  \label{intert}
\end{equation}
where $H_m$ is the differential operator
\begin{equation}
H_m:=i\gamma^0\gamma^1\frac{\partial}{\partial x}+
\frac{i}{\sin x}\left[\gamma^0\gamma^2\left(\frac{\partial}{\partial \theta}
+\frac{1}{2\tan \theta} \right)
+\frac{1}{\sin \theta}\gamma^0\gamma^3\frac{\partial}{\partial \varphi}\right]
-\frac{m}{\cos x}\gamma^0,\;\;m=M\sqrt{\frac{3}{\Lambda}}.
  \label{H}
\end{equation}
The problem essentially consists in
finding  self-adjoint realizations of $H_m$ in ${\mathcal L}^2$.
The difficulty comes from the blow-up of the gravitational
interaction on the boundary.
We see that $H_0$ is just the Dirac operator on the 3-sphere
$S^3\leftrightarrow[0,\pi]_x\times[0,\pi]_{\theta}\times[0,2\pi[_{\varphi}$,
restricted to the upper hemisphere $S^3_+\leftrightarrow[0,\frac{\pi}{2}[_x\times[0,\pi]_{\theta}\times[0,2\pi[_{\varphi}$.
The key result, Theorem \ref{proreg}, deals with the asymptotic
behaviour of $\Phi\in D(H_m)$ at the equatorial 2-sphere
$S^2=\partial S^3_+$,
as $x\rightarrow\frac{\pi}{2}$. The tool is a careful analysis based
on the diagonalization of $D_{S^2}$ by the spinoidal spherical harmonics.

%%%%%%%%%%%%%%%%%%%%%  DECOMPOSITION
%%%%%%%%%%%%%%%%%%%%%
%%%%%%%%%%%%%%%%%%%%%%%%%%%%%%%%%%%%%%%%%%%%%%%%%%%%%%%%%%%%%%%%%%%%%%%%%%%%%%
%%%%%%%%%%%%%%%%%%%%%%%%%%%%%%%%%%%%%%%%%%%%%%%%%%%%%%%%%%%%%%%%%%%%%%%%%%%%%%%%

\section{The spinoidal spherical harmonics}

We start by introducing several tools based on the spinor representation of
the rotation group (see \cite{gel}, \cite{JPN-Dirac}, \cite{vil}).
It is well known  that
there exists two Hilbert bases of $L^2(S^2)$, given by :
$\left(T^l_{\frac{1}{2},n}(\theta,\varphi)\right)_{(l,n)\in I}$,
 $\left(T^l_{-\frac{1}{2},n}(\theta,\varphi)\right)_{(l,n)\in I}$
\begin{equation}
I:=\left\{(l,n);\;l\in\NN +\frac{1}{2},\;\; n\in
  \ZZ+\frac{1}{2},\;\;l-\mid n \mid \in \NN\right\}
=\left\{(l,n);\;l\in\NN +\frac{1}{2},\;\; n=-l,-l+1,...,l\right\},
\label{l,n}
\end{equation}

$$
T^l_{\pm\frac{1}{2},n}(\theta,\varphi)=e^{ -i n\varphi}P^l_{\pm\frac{1}{2},n}(\cos\theta),
 $$
where $P^l_{\pm\frac{1}{2},n}$ can be expressed in terms of generalized Jacobi
functions :
$$
P^l_{\pm\frac{1}{2},n}(X)=A^l_{\pm,n}(1-X)^{\frac{\pm 1-2n}{4}}(1+X)^{\frac{\mp
  1-2n}{4}}\frac{d^{l-n}}{dX^{l-n}}
\left[(1-X)^{l\mp\frac{1}{2}}(1+X)^{l\pm\frac{1}{2}}\right],
  $$
and the constant
$$
A^l_{\pm,n}=\frac{(-1)^{l\mp\frac{1}{2}}i^{n\mp\frac{1}{2}}}{2^l\left(l\mp\frac{1}{2}\right)!}\sqrt{\frac{
\left(l\mp\frac{1}{2}\right)!(l+n)!}{\left(l\pm\frac{1}{2}\right)!(l-n)!}}\sqrt{\frac{2l+1}{4\pi}}
 $$
is choosen to normalize the basis functions (in comparison with the
notations  adopted in the book \cite{gel}, the functions $P^l_{m,n}$
are multiplied by $\sqrt{(2l+1)/4\pi}$)
:
$$
\int_0^{2\pi}\int_0^{\pi}T^l_{\pm\frac{1}{2},n}(\theta,\varphi)\overline{T^{l'}_{\pm\frac{1}{2},n'}(\theta,\varphi)}\sin\theta
d\theta d\varphi=\delta_{l,l'}\delta_{n,n'}.
$$
Therefore we can expand any function $f\in L^2(S^2)$ on these both
bases 
$$
f(\theta,\varphi)=\sum_{(l,n)\in I}u^l_{\pm,n}(f)T^l_{\pm\frac{1}{2},n}(\theta,\varphi),\;\;u^l_{\pm,n}(f)\in\CC,
$$
and by the Plancherel formula :
$$
\parallel f\parallel_{L^2}^2=\sum_{(l,n)\in I}\mid
u^l_{+,n}(f)\mid^2=\sum_{(l,n)\in I}\mid u^l_{-,n}(f)\mid^2.
$$
More generally, for $s\in\RR$, we introduce the Hilbert spaces
$W_{\pm}^s$ defined as the closure of the space
\begin{equation}
W_f^{\pm}:=\left\{\sum_{finite}u^l_{\pm,n}T^l_{\pm\frac{1}{2},n};\;\;u^l_{\pm,n}\in\CC\right\}
  \label{wop}
\end{equation}
 for the norm
$$
\parallel f\parallel_{W_{\pm}^s}^2:=\sum_{(l,n)\in I}\left(l+\frac{1}{2}\right)^{2s}\mid
u^l_{\pm,n}(f)\mid^2.
 $$
We note that  the basis functions are {\it not} smooth
on $S^2$ since
$T^l_{\pm\frac{1}{2},n}(\theta,2\pi)=-T^l_{\pm\frac{1}{2},n}(\theta,0)\xout\equiv\,
 0$.
Hence $W_{\pm}^s$ is not a classical Sobolev space on $S^2$. We state
some properties of these spaces. Firstly it is easy to prove that for
$$
s\geq 0\Longrightarrow W_{\pm}^s=\left\{f\in
  L^2\left(S^2\right);\;\;\parallel f\parallel_{W^s_{\pm}}<\infty\right\},
$$
and the topological dual of $W_{\pm}^s$ can be isometrically
identified with  $W_{\pm}^{-s}$
:
$$
s\in\RR,\;\;\left(W_{\pm}^s\right)'=W_{\pm}^{-s}.
$$

Secondly we show that $W_{\pm}^s$ contains the test functions on
$]0,\pi[_{\theta}\times ]0,2\pi[_{\varphi}$.  To see that, we recall the
differential equations satisfied by the basis functions :

\begin{equation}
\left(\frac{\partial}{\partial \theta}
+\frac{1}{2\tan \theta} \right)T^l_{\pm\frac{1}{2},n}=\pm\frac{n}{\sin\theta}
T^l_{\pm\frac{1}{2},n}-i\left(l+\frac{1}{2}\right)T^l_{\mp\frac{1}{2},n},
\label{equT}
\end{equation}
\begin{equation}
\frac{\partial}{\partial \varphi}T^l_{\pm\frac{1}{2},n}=
-inT^l_{\pm\frac{1}{2},n}.
\end{equation}
If $f\in
C^{\infty}_0(]0,\pi[_{\theta}\times ]0,2\pi[_{\varphi})$ then $(\partial_{\theta}+\frac{1}{2}\cot\theta\mp\frac{i}{\sin\theta}\partial_{\varphi})f\in
C^{\infty}_0(]0,\pi[_{\theta}\times ]0,2\pi[_{\varphi})$ and for any integer $N$, the
differential equation (\ref{equT}) assures that
$$
\left(l+\frac{1}{2}\right)^{2N}u_{\pm,n}^l(f)=(-1)^Nu_{\pm,n}^l\left(\left[\frac{\partial}{\partial \theta}
+\frac{1}{2\tan \theta}
\mp\frac{i}{\sin\theta}\frac{\partial}{\partial\varphi}\right]^{2N}f\right)\in
l^2(I).
$$
We conclude that any test function belongs to $W^s_{\pm}$ for any real $s$, and the
series $\Sigma_Iu^l_{\pm,n}T^l_{\pm\frac{1}{2},n}\in W^s_{\pm}$
converges in the sense of the distributions on
$]0,\pi[_{\theta}\times ]0,2\pi[_{\varphi}$, in particular for all
$s<0$. We deduce that
$(\partial_{\theta}+\frac{1}{2}\cot\theta\mp\frac{i}{\sin\theta}\partial_{\varphi})$,
acting in the sense of the distributions,
is an isometry from $W_{\pm}^s$ onto $W^{s-1}_{\mp}$.
But we have to be carefull since the set of the test functions is not dense in general in
$W_{\pm}^s$, $s>0$ :  we cannot identify $W^{-s}$ with a subspace of
distributions, and there can exist $f\in W_{\pm}^{-s}\setminus\{0\}$
which is null in the sense of the distributions on
$]0,\pi[_{\theta}\times]0,2\pi[_{\varphi}$. For instance, since
$(\sin\theta)^{-\frac{1}{2}}\in L^2(S^2)$, we have
$$
f_{\pm}:=\sum_{(l,n)\in
  I}\left(l+\frac{1}{2}\right)u_{\mp,n}^l\left(\frac{1}{\sqrt{\sin
      \theta}}\right)T^l_{\pm\frac{1}{2},n}\in
W_{\pm}^{-1},\;\;\parallel f_{\pm}\parallel_{W^{-1}_{\pm}}=\sqrt{2}\pi,
$$
but its restriction on the test functions is the null distribution
because
$$
{f_{\pm}}_{\vert C_0^{\infty}(]0,\pi[_{\theta}\times]0,2\pi[_{\varphi})}=i\left[\frac{\partial}{\partial \theta}
+\frac{1}{2\tan \theta}
\mp\frac{i}{\sin\theta}\frac{\partial}{\partial\varphi}\right]\left(\frac{1}{\sqrt{\sin\theta}}\right)=0\;\;in\;\;{\mathcal
  D}'(]0,\pi[\times]0,2\pi[).
$$

Finally we investigate the links between $W^s_+$ and $W^s_-$. We know that
$$
P^l_{\frac{1}{2},n}=P^l_{-\frac{1}{2},-n},
$$
and
$$
\overline{P^l_{\pm\frac{1}{2},n}}=(-1)^{n\mp\frac{1}{2}}P^l_{\pm\frac{1}{2},n},
$$
hence 
$$
\overline{u^l_{\pm,n}(f)}=(-1)^{n\mp\frac{1}{2}}u^l_{\mp,-n}(\overline{f}),
$$
and we have
$$
s\in\RR,\;\;\;f\in W_{\pm}^s\Longleftrightarrow\overline{f}\in
W_{\mp}^s,\;\;\parallel f\parallel_{W^s_{\pm}}=\parallel \overline{f}\parallel_{W^s_{\mp}}.
 $$
We warn that in general $W^s_+\neq W^s_-$. Indeed,
 given $f_{\pm}\in W^1_{\pm}$, we have by (\ref{equT}) :
$$
\left(\frac{\partial}{\partial \theta}
+\frac{1}{2\tan \theta}
\mp\frac{i}{\sin\theta}\frac{\partial}{\partial\varphi}\right)f_{\pm}
=\sum_{(l,n)\in
  I}-i\left(l+\frac{1}{2}\right)u_{\pm,n}^l(f_{\pm})T^l_{\mp\frac{1}{2},n}\in L^2(S^2).
$$
We deduce that 
$$
f\in W^1_+\cap W^1_-\Longrightarrow
\frac{1}{\sin\theta}\frac{\partial}{\partial\varphi}f\in L^2(S^2).
$$
Then if we consider
$$
T^{\frac{1}{2}}_{\frac{1}{2},\frac{1}{2}}(\theta,\varphi)=\sqrt{\frac{3}{4\pi}}e^{-i\frac{\varphi}{2}}\cos\frac{\theta}{2}\in
W^1_+,
$$
 we see that
$\frac{1}{\sin\theta}\partial_{\varphi}T^{\frac{1}{2}}_{\frac{1}{2},\frac{1}{2}}\notin
L^2(S^2)$, and we conclude that
$$
W^1_+\neq W^1_-.
 $$
Therefore it is convenient to introduce the isometry $\mathcal J$ on
$L^2(S^2)$ defined by
$$
{\mathcal J}\left(T_{+\frac{1}{2},n}^l\right)=T_{-\frac{1}{2},n}^l.
$$
Then we have
$$
{\mathcal J}^*\left(T_{-\frac{1}{2},n}^l\right)=T_{+\frac{1}{2},n}^l,
 $$
and $\mathcal J$ is an isometry from $W_+^s$ onto $W_-^s$.

We now return to the Dirac field. By the same way, we can expand any spinor defined on $S^2$,
$\Phi(\theta,\varphi)\in L^2(S^2;\CC^4)$ :
\begin{equation*}
\Phi(\theta,\varphi)=\sum_{(l,n)\in I}
\left(
\begin{array}{c}
u^l_{1,n}T^l_{-\frac{1}{2},n}(\theta,\varphi)\\
u^l_{2,n}T^l_{+\frac{1}{2},n}(\theta,\varphi)\\
u^l_{3,n}T^l_{-\frac{1}{2},n}(\theta,\varphi)\\
u^l_{4,n}T^l_{+\frac{1}{2},n}(\theta,\varphi)
\end{array}
\right),\;\;u^l_{j,n}\in\CC.
\end{equation*}
The main interest of this expansion is the following : if we consider the angular part of the hamiltonian $H_m$,
$$
{\mathbf D}:=i\gamma^0\gamma^2\left(\frac{\partial}{\partial \theta}
+\frac{1}{2\tan \theta} \right)
+\frac{i}{\sin \theta}\gamma^0\gamma^3\frac{\partial}{\partial \varphi},
$$
an elementary but tedious computation shows that :
\begin{equation}
{\mathbf D}\Phi(\theta,\varphi)=\sum_{(l,n)\in I}\left(l+\frac{1}{2}\right)
\left(
\begin{array}{c}
u^l_{4,n}T^l_{-\frac{1}{2},n}(\theta,\varphi)\\
u^l_{3,n}T^l_{+\frac{1}{2},n}(\theta,\varphi)\\
u^l_{2,n}T^l_{-\frac{1}{2},n}(\theta,\varphi)\\
u^l_{1,n}T^l_{+\frac{1}{2},n}(\theta,\varphi)
\end{array}
\right).
\label{decompote}
\end{equation}
Hence it is natural to  introduce the  Hilbert spaces
$$
{\mathcal W}^s:=W^s_-\times W^s_+\times W^s_-\times W^s_+
$$
endowed with the norm :
\begin{equation}
\parallel\Phi\parallel_{{\mathcal W}^s}^2:=\sum_{j=1}^4\sum_{(l,n)\in
    I}\left(l+\frac{1}{2}\right)^{2s}\mid u^l_{j,n}\mid^2.
  \label{normls}
\end{equation}
${\mathcal W}^s$ is also the closure for this norm, of the subspace
$$
{\mathcal W}_f:=W_f^-\times W_f^+\times W_f^-\times W_f^+.
 $$
As a differential operator, ${\mathbf D}$ acts from ${\mathcal W}^s$
to ${\mathcal W}^{s-1}$ and  ${\mathbf D}$ endowed with the domain
$\mathcal W^1$ is self-adjoint on $\mathcal W^0$.
We see that the spectrum of $(\mathbf D,\mathcal W^1)$ is
$\left\{\pm\left(l+\frac{1}{2}\right),\;\;l\in\NN\right\}$, its
positive subspace $L_+^2\left(S^2;\CC^4\right)$ is spanned by the
eigenvectors
$\left(T_{-\frac{1}{2},n}^l,0,0,T_{+\frac{1}{2},n}^l\right)$,
$\left(0,T_{+\frac{1}{2},n}^l,T_{-\frac{1}{2},n}^l,0\right)$,
$(l,n)\in I$, and the negative subspace $L_-^2\left(S^2;\CC^4\right)$ is spanned by the
eigenvectors
$\left(T_{-\frac{1}{2},n}^l,0,0,-T_{+\frac{1}{2},n}^l\right)$,
$\left(0,T_{+\frac{1}{2},n}^l,-T_{-\frac{1}{2},n}^l,0\right)$,
$(l,n)\in I$. We can characterize these spaces by using operator
$\mathcal J$ :
\begin{equation*}
L_{\pm}^2\left(S^2;\CC^4\right)=\left\{
\left(
\begin{array}{c}
\psi\\
\chi\\
\pm{\mathcal J}\chi\\
\pm{\mathcal J}^*\psi
\end{array}
\right),
\;\;\psi,\chi\in L^2\left(S^2\right)\right\}.
\end{equation*}
We easily obtain the orthogonal projectors $\mathbf K_{\pm}$ on
$L_{\pm}^2\left(S^2;\CC^4\right)$ :
\begin{equation}
{\mathbf K}_{\pm}=
\frac{1}{2}\left(
\begin{array}{cccc}
1&0&0&\pm\mathcal J\\
0&1&\pm{\mathcal J}^*&0\\
0&\pm \mathcal J&1&0\\
\pm{\mathcal J}^*&0&0&1
\end{array}
\right).
  \label{kaps}
\end{equation}
$\mathbf K_{\pm}$ can be extended into bounded operators on $\mathcal
W^s$, $s\in\RR$. These operators are used to define the
global boundary conditions of M.F. Atiyah, V. K. Patodi, and
I. M. Singer (see e.g. \cite{booss}) :
\begin{equation}
{\mathbf K}_{\pm}\Phi=0,
  \label{aps}
\end{equation}
and the boundary condition introduced by O. Hijazi, S. Montiel,
A. Roldan \cite{hijazi}
$$
{\mathbf K}_+\left(Id+\gamma^1\right)\Phi=0.
$$
$\mathcal W^s$ is also invariant by the operator 
$$
\mathbf B_{\alpha}:=\gamma^1+ie^{i\alpha\gamma^5},\;\;\alpha\in\RR,
$$
involved in the local MIT-bag boundary condition :
$$
\mathbf B_{0}\Phi=0,
$$
and the chiral condition :
$$
\mathbf B_{\pi}\Phi=0.
$$

If we consider the operator ${\mathbf \Lambda}:=\gamma^0\gamma^2{\mathbf D}$ as a
positive, unbounded, selfadjoint operator on ${\mathcal W}^0$ with
domain ${\mathcal W}^1$, then for $0\leq s\leq 1$, ${\mathcal W}^s$
is the domain of ${\mathbf\Lambda}^s$, that is to say, these spaces are
spaces of interpolation (see e.g. \cite{lions})  :
$$
{\mathcal W}^s=\left[{\mathcal W}^1,{\mathcal
    W}^0\right]_{1-s},\;\;0\leq s\leq 1.
$$
The link between this space and the usual Sobolev spaces on $S^2$ is
given by the following :

%%%%%%%%%%%%%%%%%%%%%%%%%%%%%%%%%%%%%%%%%%%%%%%%%%%%%%%%%%%%%%%%%%%%%
%%%%%%%%%%%%%%%%%%%%%%%%%%%%%%%%%%%%%%%%%%%%%%%%%%%%%%%%%%%%%%%%%%%%%

\begin{Proposition}
For any $s\in\RR$, the linear map
$$
{\Phi}(\theta,\varphi)\longmapsto
\Psi(x^1,x^2,x^3)=S(\theta,\varphi)\Phi(\theta,\varphi),\;\;(x^1,x^2,x^3)\in
S^2,
$$
defined from ${\mathcal W}_f$ to $\left[L^2\left(S^2\right)\right]^4$,
where $S$ is given by (\ref{SSS}) and $x^j$, $\theta$, $\varphi$ are
related to (\ref{cart}),
can be extended into a bounded isomorphism  from ${\mathcal W}^s$ onto $\left[H^s(S^2)\right]^4$.
  \label{WSHS}
\end{Proposition}

%%%%%%%%%%%%%%%%%%%%%%%%%%%%%%%%%%%%%%%%%%%%%%%%%%%%%
%%%%%%%%%%%%%%%%%%%%%%%%%%%%%%%%%%%%%%%%%%%%%%%%%%%%%

{\it Proof of Proposition \ref{WSHS}.}
A tedious but elementary calculation shows that :
\begin{equation*}
S(\theta,\varphi)=
\left(
\begin{array}{cc}
S_{11}&S_{12}\\
S_{21}&S_{22}
\end{array}
\right)
\end{equation*}
where
\begin{equation}
S_{11}=S_{22}=
\frac{1}{2}
\left(
\begin{array}{cc}
(1+i)\left(e^{-i\frac{\varphi}{2}}\cos\frac{\theta}{2}+e^{i\frac{\varphi}{2}}\sin\frac{\theta}{2}\right)&(1+i)\left(e^{i\frac{\varphi}{2}}\cos\frac{\theta}{2}-e^{-i\frac{\varphi}{2}}\sin\frac{\theta}{2}\right)\\
(1-i)\left(-e^{-i\frac{\varphi}{2}}\cos\frac{\theta}{2}+e^{i\frac{\varphi}{2}}\sin\frac{\theta}{2}\right)&(1-i)\left(e^{i\frac{\varphi}{2}}\cos\frac{\theta}{2}+e^{-i\frac{\varphi}{2}}\sin\frac{\theta}{2}\right)
\end{array}
\right),
  \label{sunun}
\end{equation}
\begin{equation*}
S_{12}=S_{21}=
\left(
\begin{array}{cc}
0&0\\
0&0
\end{array}
\right).
\end{equation*}

Following \cite{vilk}, p.337, formula (3) with $n=0$, we have
$$
\sqrt{l+1}P_{m-\frac{1}{2},-\frac{1}{2}}^{l+\frac{1}{2}}(\cos\theta)=
\sqrt{l-m+1}\cos\left(\frac{\theta}{2}\right)P_{m,0}^l(\cos\theta)
+\sqrt{l+m}\sin\left(\frac{\theta}{2}\right)P_{m-1,0}^l(\cos\theta),
$$
then since
$$
P^l_{m,n}=(-1)^{m+n}P^l_{n,m},
$$
we get for $l\in\NN$, $m\in\ZZ$, $-l\leq m\leq l+1$ :
\begin{equation*}
\begin{split}
e^{-i\frac{\varphi}{2}}\cos\left(\frac{\theta}{2}\right)T_{-\frac{1}{2},m-\frac{1}{2}}^{l+\frac{1}{2}}&(\theta,\varphi)=\\
&(-1)^{m-1}\sqrt{\frac{l-m+1}{l+1}}\frac{x^3+1}{2}Y^l_m(\theta,\varphi)
+
(-1)^{m-1}\sqrt{\frac{l+m}{l+1}}\frac{x^1-ix^2}{2}Y^l_{m-1}(\theta,\varphi)
\end{split}
\end{equation*}
\begin{equation*}
\begin{split}
e^{i\frac{\varphi}{2}}\sin\left(\frac{\theta}{2}\right)T_{-\frac{1}{2},m-\frac{1}{2}}^{l+\frac{1}{2}}&(\theta,\varphi)=\\
&(-1)^{m-1}\sqrt{\frac{l-m+1}{l+1}}\frac{x^1+ix^2}{2}Y^l_m(\theta,\varphi)
+
(-1)^{m-1}\sqrt{\frac{l+m}{l+1}}\frac{1-x^3}{2}Y^l_{m-1}(\theta,\varphi)
\end{split}
\end{equation*}
By the same way, with \cite{vilk}, p.337, formula (4) with $n=0$, we have
\begin{equation*}
\sqrt{l+1}P_{m-\frac{1}{2},\frac{1}{2}}^{l+\frac{1}{2}}(\cos\theta)=
-\sqrt{l-m+1}\sin\left(\frac{\theta}{2}\right)P_{m,0}^l(\cos\theta)
+\sqrt{l+m}\cos\left(\frac{\theta}{2}\right)P_{m-1,0}^l(\cos\theta),
\end{equation*}
then 
we get for $l\in\NN$, $m\in\ZZ$, $-l+1\leq m\leq l$ :
\begin{equation*}
\begin{split}
e^{i\frac{\varphi}{2}}\cos\left(\frac{\theta}{2}\right)T_{\frac{1}{2},m-\frac{1}{2}}^{l+\frac{1}{2}}&(\theta,\varphi)=\\
&(-1)^{m+1}\sqrt{\frac{l-m+1}{l+1}}\frac{x^1+ix^2}{2}Y^l_m(\theta,\varphi)
+
(-1)^{m}\sqrt{\frac{l+m}{l+1}}\frac{x^3+1}{2}Y^l_{m-1}(\theta,\varphi),
\end{split}
\end{equation*}
\begin{equation*}
\begin{split}
e^{-i\frac{\varphi}{2}}\sin\left(\frac{\theta}{2}\right)T_{\frac{1}{2},m-\frac{1}{2}}^{l+\frac{1}{2}}&(\theta,\varphi)=\\
&(-1)^{m+1}\sqrt{\frac{l-m+1}{l+1}}\frac{1-x^3}{2}Y^l_m(\theta,\varphi)
+
(-1)^{m}\sqrt{\frac{l+m}{l+1}}\frac{x^1-ix^2}{2}Y^l_{m-1}(\theta,\varphi).
\end{split}
\end{equation*}
Since $f\mapsto x^jf$ is bounded on $H^s(S^2)$ and 
$f$ belongs to $H^s(S^2)$ iff
$$
f=\sum_{l=0}^{\infty}\sum_{m=-l}^l\alpha_{l,m}Y^l_m,\;\;\sum_{l,m}l^{2s}\mid\alpha_{l,m}\mid^2<\infty,
$$
we conclude that the linear map $\Phi\mapsto S(\theta,\varphi)\Phi=\Psi$ is bounded
from ${\mathcal W}_f$ endowed with the norm of ${\mathcal W}^s$ to
$\left[H^s(S^2)\right]^4$, hence it can be extended into a continuous
linear map $\mathbb S:\Phi\mapsto\Psi$ from ${\mathcal W}^s$ to
$\left[H^s(S^2)\right]^4$. Then $\mathbb S^*$ is a bounded linear map
from $\left[H^{-s}(S^2)\right]^4$ to ${\mathcal W}^{-s}$ for any $s\in
\RR$. Since
${\mathbb S^*}\Psi=S^*(\theta,\varphi)\Psi$ for $\Psi\in
\left[C^{\infty}_0(S^2)\right]^4$, and
$S^*(\theta,\varphi)=S^{-1}(\theta,\varphi)$, we conclude that
$\mathbb S\mathbb S^*=Id_{H^s}$, $\mathbb S^*\mathbb S= Id_{{\mathcal
    W}^s}$.

\fin

%%%%%%%%%%%%%%%%%%%%%%%%%%%%%%%%%%%%%%%%%%%%%%%%%%%%%%%%%%%%%%%%%%%%%%%%%%%%%%%%%%%%%%%
%%%%%%%%%%%%%%%%%%%%%%%%%%%%%%%%%%%%%%%%%%%%%%%%%%%%%%%%%%%%%%%%%%%%%%%%%%%%%%%%%%%%%%%%%
\section{Asymptotic behaviour at the boundary}

In this part we investigate  the properties of the spinors that belong to
the natural domain of the hamiltonian, especially the asymptotic behaviours near the
boundary. We begin with its form
in spherical coordinates, $H_m$ given by (\ref{H}), and 
\begin{equation}
D\left(H_m\right):=\left\{\Phi\in{\mathcal L}^2;\;\; H_m\Phi\in{\mathcal L}^2\right\}.
  \label{dhm}
\end{equation}

%%%%%%%%%%%%%%%%%%%%%%%%%%%%%%%%%%%%%%%%%%%%%%%%%%%%%%%%%%%%%%%%%%%%%%%%%%%%%%%%%%%%%%%%%%%%%%%%
%%%%%%%%%%%%%%%%%%%%%%%%%%%%%%%%%%%%%%%%%%%%%%%%%%%%%%%%%%%%%%%%%%%%%%%%%%%%%%%%%%%%%%%%%%%%%%%%

\begin{Theorem}
For any $\Phi\in D\left(H_m\right)$  we
have
\begin{equation}
\Phi\in C^0\left([0,\frac{\pi}{2}[_x;{\mathcal W}^{\frac{1}{2}}\right),
  \label{regfi}
\end{equation}
\begin{equation}
\parallel \Phi(x,.)\parallel_{{\mathcal W}^{\frac{1}{2}}}=O\left(\sqrt x\right),\;\;x\rightarrow 0,
  \label{compo}
\end{equation}
and when $0<m$ we have
\begin{equation}
\int_0^{\frac{\pi}{2}}\parallel \Phi(x,.)\parallel^2_{{\mathcal
    W}^1}\frac{dx}{\sin x}\leq
\parallel H_m\Phi\parallel_{{\mathcal L}^2}^2.
  \label{estow}
\end{equation}
When
$
\frac{1}{2}< m$,
we have
\begin{equation}
\parallel \Phi(x,.)\parallel_{L^2(S^2)}=O\left(\sqrt{\frac{\pi}{2}-x}\right),\;\;x\rightarrow \frac{\pi}{2}.
  \label{deko}
\end{equation}
When
$ m=\frac{1}{2}$,
we have :
\begin{equation}
\parallel \Phi(x,.)\parallel_{L^2(S^2)}=O\left(\sqrt{\left(x-\frac{\pi}{2}\right)\ln\left(\frac{\pi}{2}-x\right)}\right),\;\;x\rightarrow \frac{\pi}{2}.
  \label{deku}
\end{equation}
When 
$
0<m<\frac{1}{2}$,
there exists $\psi_-\in W_-^{\frac{1}{2}}$,
$\chi_-\in W_+^{\frac{1}{2}}$, $\psi_{+},\chi_{+}\in L^2(S^2)$, and $\phi\in
C^0\left([0,\frac{\pi}{2}]_x;L^2(S^2;\CC^4)\right)$
satisfying
\begin{equation}
\Phi(x,\theta,\varphi)=
\left(\frac{\pi}{2}-x\right)^{-m}
\left(
\begin{array}{c}
\psi_-(\theta,\varphi)\\
\chi_-(\theta,\varphi)\\
-i\psi_-(\theta,\varphi)\\
i\chi_-(\theta,\varphi)
\end{array}
\right)
+
\left(\frac{\pi}{2}-x\right)^{m}
\left(
\begin{array}{c}
\psi_+(\theta,\varphi)\\
\chi_+(\theta,\varphi)\\
i\psi_+(\theta,\varphi)\\
-i\chi_+(\theta,\varphi)
\end{array}
\right)+\phi(x,\theta,\varphi),
\label{dqf}
\end{equation}
\begin{equation}
\parallel \phi(x,.)\parallel_{L^2(S^2)}=o\left(\sqrt{\frac{\pi}{2}-x}\right),\;\;x\rightarrow \frac{\pi}{2}.
  \label{faq}
\end{equation}
Conversely, for any  $\psi_-\in W_-^{\frac{1}{2}+m}$,
$\chi_-\in W_+^{\frac{1}{2}+m}$, $\psi_{+}\in W_-^{\frac{1}{2}-m}$,
$\chi_{+}\in W_+^{\frac{1}{2}-m}$ there exists $\Phi \in D(H_m)$
satisfying (\ref{dqf}) and (\ref{faq}).\\

When
$
m=0$, then
\begin{equation}
\Phi\in C^0\left([0,\frac{\pi}{2}]_x;{\mathcal W}^{-\frac{1}{2}}\right).
  \label{regmo}
\end{equation}

  \label{proreg}
\end{Theorem}

%%%%%%%%%%%%%%%%%%%%%%%%%%%%%%%%%%%%%%%%%%%%%%%%%%%%%%%%%%%%%%%%%%%%%%%%%%%%%%%%%%%%%%%
%%%%%%%%%%%%%%%%%%%%%%%%%%%%%%%%%%%%%%%%%%%%%%%%%%%%%%%%%%%%%%%%%%%%%%%%%%%%%%%%%%%%%%%%%

\begin{Remark}
(\ref{estow}) shows that when $m>0$, $H_m\Phi=0$ implies $\Phi=0$.
In opposite, when  $m=0$, the
left member of (\ref{estow}) can be infinite even if
$H_0\Phi=0$. Furthermore the space ${\mathcal W}^{-\frac{1}{2}}$ is
optimal for the traces on $x=\frac{\pi}{2}$ : there exists $\Phi\in
D(H_0)$ such that
$\Phi(\frac{\pi}{2})\notin\cup_{s>-\frac{1}{2}}{\mathcal W}^s$. As an
example, we consider a
sequence $\left(C_{l,n}\right)_{(l,n)\in I}\subset\CC$ such that
$$
\sum_{(l,n)\in I}\left(l+\frac{1}{2}\right)^{-1}\left\vert C_{l,n}\right\vert^2<\infty,\;\;
-1<s\Rightarrow\sum_{(l,n)\in I}\left(l+\frac{1}{2}\right)^{s}\left\vert C_{l,n}\right\vert^2=\infty,
$$
we can take for instance $C_{l,n}=\frac{1}{\sqrt{l}\log(l+1)}$, and we put
$$
\Phi(x,\theta,\varphi)=\sum_{(l,n)\in I}C_{l,n}\tan\left(\frac{x}{2}\right)^{l+\frac{1}{2}}
\left(\begin{array}{c}
T^l_{-\frac{1}{2},n}(\theta,\varphi)\\
-iT^l_{\frac{1}{2},n}(\theta,\varphi)\\
0\\
0
\end{array}
\right).
$$
Then we easily check that
$$
\Phi\in{\mathcal
  L}^2,\;\; H_0\Phi=0,\;\;\;0<s\Rightarrow\int_0^{\frac{\pi}{2}}\parallel
\Phi(x,.)\parallel_{{\mathcal
    W}^s}^2dx=\infty,\;\;\Phi(\frac{\pi}{2},.)\in
{\mathcal W}^{-\frac{1}{2}}\setminus\cup_{s>-\frac{1}{2}}{\mathcal W}^s.
$$
  \label{remb}
\end{Remark}
\begin{Remark}
For $0<m<\frac{1}{2}$, the leading terms of $\Phi$ satisfy the
MIT-bag or the Chiral boundary condition since :
\begin{equation*}
{\mathbf B}_0\Phi(x)=
2i
\left(\frac{\pi}{2}-x\right)^{m}
\left(
\begin{array}{c}
\psi_+\\
\chi_+\\
i\psi_+\\
-i\chi_+
\end{array}
\right)+{\mathbf B}_0\phi(x),
\;\;
{\mathbf B}_{\pi}\Phi(x)=
-2i
\left(\frac{\pi}{2}-x\right)^{-m}
\left(
\begin{array}{c}
\psi_-\\
\chi_-\\
-i\psi_-\\
i\chi_-
\end{array}
\right)+{\mathbf B}_{\pi}\phi(x).
\end{equation*}
  \label{remit}
\end{Remark}

%%%%%%%%%%%%%%%%%%%%%%%%%%%%%%%%%%%%%%%%%%%%%%%%%%%%%%%%%%%%%%%%%%%%%%%%%%%%%%%%%%%%%%
%%%%%%%%%%%%%%%%%%%%%%%%%%%%%%%%%%%%%%%%%%%%%%%%%%%%%%%%%%%%%%%%%%%%%%%%%%%%%%%%%%%%%%%%

%%%%%%%%%%%%%%%%%%%%%%%%%%%%%%%%%%%%%%%%%%%%%%%%%%%%%%%%%%%%%%%%%%%%%%%%%%%%%%%%%%%%%%%%%%%%%%%%%%%%%%%%%%%%%%%%%%%%%%%%%%%%%%%%%%%%%%%%%%%%%%%%%%%%%%%%%%%%%%%%%%%%%%%%%%%%%%%%%%%%%%%%%%%%%%%%%%

{\it Proof of Theorem \ref{proreg}.} We  expand any spinor $\Phi(x,\theta,\varphi)$ by the
previous way :
\begin{equation*}
\Phi(x,\theta,\varphi)=\sum_{(l,n)\in I}
\left(
\begin{array}{c}
u^l_{1,n}(x)T^l_{-\frac{1}{2},n}(\theta,\varphi)\\
u^l_{2,n}(x)T^l_{+\frac{1}{2},n}(\theta,\varphi)\\
u^l_{3,n}(x)T^l_{-\frac{1}{2},n}(\theta,\varphi)\\
u^l_{4,n}(x)T^l_{+\frac{1}{2},n}(\theta,\varphi)
\end{array}
\right),
\end{equation*}
and we have :
$$
\parallel\Phi\parallel_{{\mathcal L}^2}^2=\sum_{j=1}^4\sum_{(l,n)\in
  I}\parallel u^l_{j,n}\parallel_{L^2(0,\frac{\pi}{2})}^2.
$$
Furthermore, for $\Phi\in D(H_m)$, (\ref{decompote}) gives :
\begin{equation*}
H_m\Phi(x,\theta,\varphi)=\sum_{(l,n)\in I}
\left(
\begin{array}{c}
f^l_{1,n}(x)T^l_{-\frac{1}{2},n}(\theta,\varphi)\\
f^l_{2,n}(x)T^l_{+\frac{1}{2},n}(\theta,\varphi)\\
f^l_{3,n}(x)T^l_{-\frac{1}{2},n}(\theta,\varphi)\\
f^l_{4,n}(x)T^l_{+\frac{1}{2},n}(\theta,\varphi)
\end{array}
\right),
\end{equation*}
with
\begin{equation}
\left\{
\begin{array}{c}
i\left(u^l_{3,n}\right)'+\frac{\left(l+\frac{1}{2}\right)}{\sin
    x}u^l_{4,n}-\frac{m}{\cos x}u^l_{1,n}=f^l_{1,n},\\
-i\left(u^l_{4,n}\right)'+\frac{\left(l+\frac{1}{2}\right)}{\sin
    x}u^l_{3,n}-\frac{m}{\cos x}u^l_{2,n}=f^l_{2,n},\\
i\left(u^l_{1,n}\right)'+\frac{\left(l+\frac{1}{2}\right)}{\sin
    x}u^l_{2,n}+\frac{m}{\cos x}u^l_{3,n}=f^l_{3,n},\\
-i\left(u^l_{2,n}\right)'+\frac{\left(l+\frac{1}{2}\right)}{\sin
    x}u^l_{1,n}+\frac{m}{\cos x}u^l_{4,n}=f^l_{4,n},
\end{array}
\right.,
  \label{sistu}
\end{equation}
and
$$
\parallel H_m\Phi\parallel_{{\mathcal L}^2}^2=\sum_{j=1}^4\sum_{(l,n)\in
  I}\parallel f^l_{j,n}\parallel_{L^2(0,\frac{\pi}{2})}^2.
$$

For $1\leq h,k\leq4$, we put
$$
u^{l,\pm}_{hk,n}=u^l_{h,n}\pm iu^l_{k,n},\;\;f^{l,\pm}_{hk,n}=f^l_{h,n}\pm if^l_{k,n}.
 $$
We have
$$
\left(u^{l,\pm}_{12,n}\right)'\mp\frac{l+\frac{1}{2}}{\sin
  x}u^{l,\pm}_{12,n}=\frac{im}{\cos x}u^{l,\mp}_{34,n}-if^{l,\mp}_{34,n},
$$
$$
\left(u^{l,\pm}_{34,n}\right)'\mp\frac{l+\frac{1}{2}}{\sin
  x}u^{l,\pm}_{34,n}=-\frac{im}{\cos x}u^{l,\mp}_{12,n}-if^{l,\mp}_{12,n}.
$$
Given $w_+^l\in L^2(0,\frac{\pi}{2})$, any solution $v_+^l$ of
$$
\frac{d}{dx}v^l_+-\frac{l+\frac{1}{2}}{\sin x}v_+^l=w_+^l,\;\;0<x<\frac{\pi}{2},
$$
belongs to $H^1_{loc}(]0,\frac{\pi}{2}])\subset
C^0(]0,\frac{\pi}{2}])$ and $v_+^l$
can be written :
\begin{equation}
v_+^l(x)=v_+^l(\frac{\pi}{2})\left(\tan\left(\frac{x}{2}\right)\right)^{l+\frac{1}{2}}-\int_x^{\frac{\pi}{2}}\left(\frac{\tan\left(\frac{x}{2}\right)}{\tan\left(\frac{y}{2}\right)}\right)^{l+\frac{1}{2}}w_+^l(y)dy.
  \label{eqvpl}
\end{equation}
On the one hand, by integrating we get :
\begin{equation}
\mid v_+^l(\frac{\pi}{2})\mid^2\leq C(l+1)(\parallel
v_+^l\parallel_{L^2}^2+\parallel w_+^l\parallel_{L^2}^2).
  \label{vpi}
\end{equation}
On the other hand, we easily show that for $0<x\leq \frac{\pi}{2}$
$$
\int_x^{\frac{\pi}{2}}\left(\tan\left(\frac{y}{2}\right)\right)^{-2l-1}dy\leq
\frac{1}{2l}\left(\tan\left(\frac{x}{2}\right)\right)^{-2l}\left(1-\left(\tan\left(\frac{x}{2}\right)\right)^{2l}\right),
$$
therefore  since $\tan(x/2)\leq x$ on $[0,\frac{\pi}{2}]$, we obtain
that :
$$
2l\left\vert v_+^l(x)-v_+^l\left(\frac{\pi}{2}\right)\left(\tan\left(\frac{x}{2}\right)\right)^{l+\frac{1}{2}}\right\vert^2\leq \mid x\mid \parallel w_+^l\parallel_{L^2(x,\frac{\pi}{2})}^2\left(1-\left(\tan\left(\frac{x}{2}\right)\right)^{2l}\right).
$$
and we conclude that 
\begin{equation}
v_+^l(\frac{\pi}{2})=0\Longrightarrow l\mid v_+^l(x)\mid^2\leq \mid
x\mid \parallel w_+^l\parallel_{L^2}^2.
  \label{vpo}
\end{equation}
Now the solutions $v_-^l$  of
\begin{equation}
\frac{d}{dx}v^l_-+\frac{l+\frac{1}{2}}{\sin x}v_-^l=w_-^l\in L^2(0,\frac{\pi}{2}),
  \label{eqvmoins}
\end{equation}
have the form
\begin{equation}
v_-^l(x)=C\left(\tan\left(\frac{x}{2}\right)\right)^{-l-\frac{1}{2}}+
\int_0^x\left(\frac{\tan\left(\frac{y}{2}\right)}{\tan\left(\frac{x}{2}\right)}\right)^{l+\frac{1}{2}}w_-^l(y)dy.
  \label{equvpl}
\end{equation}
Then, when  $v_-\in L^2(0,\frac{\pi}{2})$ and $l\geq
0$, we have $C=0$. Since for $0\leq x\leq \frac{\pi}{2}$ we have
$$
\int_0^x\left(\tan\left(\frac{y}{2}\right)\right)^{2l+1}dy\leq \frac{1}{l+1}\left(\tan\left(\frac{x}{2}\right)\right)^{2l+2},
$$
we obtain that the $L^2$ solutions of (\ref{eqvmoins}) satisfy :
\begin{equation}
(l+1)\mid v_-^l(x)\mid^2\leq \mid x\mid \parallel w_-^l\parallel_{L^2}^2.
  \label{vml}
\end{equation}

For any $\chi\in C^{\infty}_0([0,\frac{\pi}{2}[)$, we apply the previous
estimates to 
$$
v_{\pm}^l=\chi u^{l,\pm}_{12(34),n},\;\;w_{\pm}^l=+(-)\frac{im}{\cos
  x}\chi u_{34(12),n}^{l,\mp}-i\chi
f_{34(12),n}^{l,\mp}-\chi'u_{12(34),n}^{l,\pm}.
$$
From (\ref{vpo}) and (\ref{vml}),  we deduce
\begin{equation}
l\sum_{hk=12,34}\mid \chi(x)u^{l,\pm}_{hk,n}(x)\mid^2 \leq C(\chi)\mid
x\mid\sum_{j=1}^4
\parallel u_{j,n}^{l}\parallel_{L^2}^2+\parallel f_{j,n}^{l}\parallel_{L^2}^2,
  \label{inepte}
\end{equation}
where $C(\chi)>0$ depends only of $\chi$. We get (\ref{regfi}) and
(\ref{compo}) that
are consequences of (\ref{inepte}). When $m=0$, we can take
$$
v_{\pm}^l=u^{l,\pm}_{12(34),n},\;\;w_{\pm}^l=-i
f_{34(12),n}^{l,\mp},
$$
and we get from (\ref{vpi}) and (\ref{vml}) that
\begin{equation}
(l+1)^{-1}\sum_{hk=12,34}\mid u^{l,\pm}_{hk,n}(x)\mid^2 \leq C\mid
x\mid\sum_{j=1}^4
\parallel u_{j,n}^{l}\parallel_{L^2}^2+\parallel f_{j,n}^{l}\parallel_{L^2}^2.
  \label{inepte'}
\end{equation}
This estimate yields to (\ref{regmo}).\\

Now we have
$$
\left(u_{13,n}^{l,\pm}\right)'\mp\frac{m}{\cos x}u_{13,n}^{l,\pm}=
\pm f_{13,n}^{l,\mp}+i\frac{l+\frac{1}{2}}{\sin x}u_{24,n}^{l,\pm},
$$
$$
\left(u_{24,n}^{l,\pm}\right)'\pm\frac{m}{\cos x}u_{24,n}^{l,\pm}=
\mp f_{24,n}^{l,\mp}-i\frac{l+\frac{1}{2}}{\sin x}u_{13,n}^{l,\pm},
$$

Given $m\geq 0$, $w_+^l\in L^2(0,\frac{\pi}{2})$, any solution $v_+^l$ of
$$
\frac{d}{dx}v^l_++\frac{m}{\cos x}v_+^l=w_+^l,\;\;0<x<\frac{\pi}{2},
$$
belongs to $H^1_{loc}([0,\frac{\pi}{2}[)\subset
C^0([0,\frac{\pi}{2}[)$ and when 
$$
v_+^l(0)=0,
$$
$v_+^l$
can be written :
\begin{equation}
v_+^l(x)=\int_0^{x}\left(\frac{\tan\left(\frac{\pi}{4}-\frac{x}{2}\right)}{\tan\left(\frac{\pi}{4}-\frac{y}{2}\right)}\right)^{m}w_+^l(y)dy.
  \label{eqppl}
\end{equation}
Therefore the Cauchy-Schwarz estimate yields
\begin{equation}
\frac{1}{2}<m\Longrightarrow \mid v_+^l(x)\mid \leq C\parallel w_+^l\parallel_{L^2}\sqrt{\frac{\pi}{2}-x},
  \label{grom}
\end{equation}
\begin{equation}
m=\frac{1}{2}\Longrightarrow \mid v_+^l(x)\mid \leq C\parallel w_+^l\parallel_{L^2}\sqrt{\left(\frac{\pi}{2}-x\right)\ln\left(\frac{\pi}{2}-x\right)},
  \label{grum}
\end{equation}
$$
0\leq m<\frac{1}{2}\Longrightarrow \mid v_+^l(x)\mid \leq C\parallel w_+^l\parallel_{L^2}\left(\frac{\pi}{2}-x\right)^m.
$$
We precise this last estimate for $0\leq m<\frac{1}{2}$ :
\begin{equation}
\begin{split}
\left\vert
  v_+^l(x)-2^{-m}\left(\frac{\pi}{2}-x\right)^m\int_0^{\frac{\pi}{2}}\left[\tan\left(\frac{\pi}{4}-\frac{y}{2}\right)\right]^{-m}w_+^l(y)dy\right\vert
\\
\leq C\left(\parallel
  w_+^l\parallel_{L^2}\left(\frac{\pi}{2}-x\right)^{m+2}+
\parallel
w_+^l\parallel_{L^2(x,\frac{\pi}{2})}\sqrt{\frac{\pi}{2}-x}\right),
\end{split}
\label{vpc}
\end{equation}
in particular we have
\begin{equation}
0<m\Longrightarrow \lim_{x\rightarrow\frac{\pi}{2}}v_+^l(x)=0.
  \label{vmo}
\end{equation}

On the other hand the solutions $v_-^l$ of
$$
\frac{d}{dx}v^l_--\frac{m}{\cos x}v_-^l=w_-^l,\;\;0<x<\frac{\pi}{2},
$$
have the form
\begin{equation}
v_-^l(x)=C_l\left[\tan\left(\frac{\pi}{4}-\frac{x}{2}\right)\right]^{-m}
-\int_x^{\frac{\pi}{2}}\left(\frac{\tan\left(\frac{\pi}{4}-\frac{y}{2}\right)}{\tan\left(\frac{\pi}{4}-\frac{x}{2}\right)}\right)^m
w_-^l(y)dy,
  \label{equpl}
\end{equation}
thus,
\begin{equation}
\left\vert
v_-^l(x)-C_l\left[\tan\left(\frac{\pi}{4}-\frac{x}{2}\right)\right]^{-m}\right\vert
\leq  \parallel w_-^l\parallel_{L^2(x,\frac{\pi}{2})}\sqrt{\frac{\pi}{2}-x},
  \label{vmc}
\end{equation}
and
\begin{equation}
v_-^l\in L^2(0,\frac{\pi}{2}),\;\;\frac{1}{2}\leq m\Longrightarrow C_l=0,
  \label{plo}
\end{equation}
$$
0\leq m<\frac{1}{2}\Longrightarrow C_l=v_-^l(0)+\int_0^{\frac{\pi}{2}}\left(\tan\left(\frac{\pi}{4}-\frac{y}{2}\right)\right)^mw_-^l(y)dy.
$$

We pick $\chi\in C^{\infty}_0(]0,\frac{\pi}{2}]$ such that
$\chi(\frac{\pi}{2})=1$, and we apply the
previous estimates to
$$
v_{\pm}^l=\chi u_{13(24),n}^{l,\mp(\pm)},\;\;w_{\pm}^l=w_{13(24),n}^{l,\mp(\pm)}:=\mp(\pm)\chi
f_{13(24),n}^{l,\pm(\mp)}+(-)i\frac{l+\frac{1}{2}}{\sin x}\chi u_{24(13),n}^{l,\mp(\pm)}-\chi'u_{13(24),n}^{l,\mp(\pm)}.
$$
From (\ref{vmo}) we deduce that when $m>0$ :
$$
\lim_{x\rightarrow\frac{\pi}{2}}u_{1,n}^l(x)-iu_{3,n}^l(x)=\lim_{x\rightarrow\frac{\pi}{2}}u_{2,n}^l(x)+iu_{4,n}^l(x)=0,
$$
hence
\begin{equation}
\lim_{x\rightarrow\frac{\pi}{2}}\Im\left(u_{1,n}^l(x)\overline{u_{2,n}^l(x)}+
(u_{3,n}^l(x)\overline{u_{4,n}^l(x)}\right)=0.
  \label{nuol}
\end{equation}
Now multiplying (\ref{sistu}) by $\overline{u_{j,n}^l}$ and taking the
real part we get :
$$
\frac{d}{dx}\Im\left(u_{1,n}^l\overline{u_{2,n}^l}+(u_{3,n}^l\overline{u_{4,n}^l}\right)+
\frac{\left(l+\frac{1}{2}\right)}{\sin x}\sum_1^4\mid u_{j,n}^l\mid^2=
\Re\left(f_{1,n}^l\overline{u_{4,n}^l}+f_{2,n}^l\overline{u_{3,n}^l}+f_{3,n}^l\overline{u_{2,n}^l}+f_{4,n}^l\overline{u_{1,n}^l}\right),
$$
and thanks to (\ref{inepte}) and (\ref{nuol}) we obtain
$$
\int_0^{\frac{\pi}{2}}\frac{\left(l+\frac{1}{2}\right)^2}{\sin
  x}\sum_{j=1}^4\mid u_{j,n}^l(x)\mid^2dx
\leq \sum_{j=1}^4\parallel f_{j,n}^l\parallel_{L^2}^2,
$$
that proves (\ref{estow}). We also see that :
\begin{equation}
\parallel w_{13(24),n}^{l,\mp(\pm)}\parallel_{L^2}\leq
C(\chi)\sum_{j=1}^4\parallel f_{j,n}^l\parallel_{L^2}.
  \label{wl}
\end{equation}
Therefore when $m\geq\frac{1}{2}$, (\ref{deko}) and (\ref{deku}) follow from (\ref{grom}),
(\ref{grum}), (\ref{vmc}) and (\ref{plo}). On the other hand, when $0<m<\frac{1}{2}$,
 (\ref{vpc}), (\ref{vmc}) and (\ref{wl}) assure there exists
$\varphi_{13(24),n}^{l,\mp(\pm}\in C^0([0,\frac{\pi}{2}])$ such that :
$$
u_{13(24),n}^{l,-(+)}(x)=\left(\frac{\pi}{2}-x\right)^m\int_0^{\frac{\pi}{2}}\left(2\tan\left(\frac{\pi}{4}-\frac{y}{2}\right)\right)^{-m}w_{13(24),n}^{l,-(+)}(y)dy
+\varphi_{13(24),n}^{l,-(+)}(x)\sqrt{\frac{\pi}{2}-x},
$$
$$
u_{13(24),n}^{l,+(-)}(x)=
\left(\frac{\pi}{2}-x\right)^{-m}\int_0^{\frac{\pi}{2}}\left(\frac{\frac{\pi}{2}-x}{\tan\left(\frac{\pi}{4}-\frac{x}{2}\right)}\tan\left(\frac{\pi}{4}-\frac{y}{2}\right)\right)^{m}w_{13(24),n}^{l,+(-)}(y)dy
+\varphi_{13(24),n}^{l,+(-)}(x)\sqrt{\frac{\pi}{2}-x},
 $$
$$
\lim_{x\rightarrow\frac{x}{2}}\sum_{(l,n)\in I}\left\vert \varphi_{13(24),n}^{l,\mp(\pm)}(x)\right\vert^2=0.
$$
We deduce that there exists $\psi_{\pm}, \chi_{\pm}\in L^2(S^2)$ such
that $\Phi$ can be expressed according to (\ref{dqf}),
(\ref{faq}). It remains to prove the regularity of $\psi_-$ and
$\chi_-$. We consider 
$$
\Psi(x,\theta,\varphi):=\left(\frac{\cos x}{1+\sin
    x}\right)^m\left(1+i\gamma^1\right)\Phi.
$$
(\ref{dqf}),
(\ref{faq}) assure that
\begin{equation}
\Psi(x,.)\longrightarrow
\left(
\begin{array}{c}
\psi_-\\
\chi_-\\
-i\psi_-\\
i\chi_-
\end{array}
\right)\;\;in\;\;{\mathcal W}^0\;\;as\;\;x\rightarrow\frac{\pi}{2}.
  \label{lif}
\end{equation}
We calculate
$$
\frac{\partial}{\partial x}\Psi(x,.)=
\left(\frac{\cos x}{1+\sin
    x}\right)^m\left(1+i\gamma^1\right)\gamma^0\left(H_m\Phi-\frac{1}{\sin x}{\mathbf D}\Phi\right).
$$
Since $\Phi\in L^2\left([0,\frac{\pi}{2}[_x;{\mathcal
    W}^1\right)$ by (\ref{estow}), we deduce that
$$
\Psi\in
L^2\left([1,\frac{\pi}{2}[_x;{\mathcal W}^1\right).
$$
$$
\frac{\partial}{\partial x}\Psi\in
L^2\left([1,\frac{\pi}{2}[_x;{\mathcal W}^0\right).
$$
The theorem of the intermediate derivative (\cite{lions}, p. 23) shows that
$$
\Psi\in C^0\left([1,\frac{\pi}{2}]_x;\left[{\mathcal W}^{1},{\mathcal W}^0\right]_{\frac{1}{2}}\right).
$$
Recalling that $\left[{\mathcal W}^{1},{\mathcal
    W}^0\right]_{\frac{1}{2}}={\mathcal W}^{\frac{1}{2}}$, we conclude
by (\ref{lif}) that $\psi_-\in W_-^{\frac{1}{2}}$, $\chi_-\in
W_+^{\frac{1}{2}}$.\\

%%%%%%%%%%%%%%%%%%%%%%%%%%%%%%%%%%%%%%%%%%%%%%%%%%%%%%%%%

Finally we consider $\psi_{\pm}\in W_-^{\frac{1}{2}\mp m}$,
$\chi_{\pm}\in W_+^{\frac{1}{2}\mp m}$, and we want to construct $\Phi\in
D(H_m)$ satisfying (\ref{dqf}) and (\ref{faq}). We choose $f\in
C^{\infty}_0([0,1[)$ such that $f(0)=1$, and we put
\begin{equation*}
\Phi(x)=
\left(\frac{\pi}{2}-x\right)^{-m}
\left(
\begin{array}{c}
\psi_-\\
\chi_-\\
-i\psi_-\\
i\chi_-
\end{array}
\right)
+
\left(\frac{\pi}{2}-x\right)^{m}
\left(
\begin{array}{c}
\psi_+\\
\chi_+\\
i\psi_+\\
-i\chi_+
\end{array}
\right)+\phi(x),
\end{equation*}
where
\begin{equation*}
\begin{split}
\phi(x)=&
\left(\frac{\pi}{2}-x\right)^{-m}\sum_{(l,n)\in I}
[f\left(l\left(\frac{\pi}{2}-x\right)\right)-1]
\left(
\begin{array}{c}
u_{-,n}^l(\psi_-)T^l_{-\frac{1}{2},n}\\
u_{+,n}^l(\chi_-)T^l_{+\frac{1}{2},n}\\
-iu_{-,n}^l(\psi_-)T^l_{-\frac{1}{2},n}\\
iu_{+,n}^l(\chi_-)T^l_{+\frac{1}{2},n}
\end{array}
\right)\\
+&
\left(\frac{\pi}{2}-x\right)^{m}\sum_{(l,n)\in I}
[f\left(l\left(\frac{\pi}{2}-x\right)\right)-1]
\left(
\begin{array}{c}
u_{-,n}^l(\psi_+)T^l_{-\frac{1}{2},n}\\
u_{+,n}^l(\chi_+)T^l_{+\frac{1}{2},n}\\
iu_{-,n}^l(\psi_+)T^l_{-\frac{1}{2},n}\\
-iu_{+,n}^l(\chi_+)T^l_{+\frac{1}{2},n}
\end{array}
\right).
\end{split}
\end{equation*}
We use the fact that
$$
\left\vert
  f\left(l\left(\frac{\pi}{2}-x\right)\right)-1\right\vert^2\leq
\left(\frac{\pi}{2}-x\right)^{1\pm
  2m}\left(\int_0^{l\left(\frac{\pi}{2}-x\right)}\mid
  f'(t)\mid^{\frac{2}{1\mp 2m}}dt\right)^{1\mp 2m}l^{1\pm 2m},
$$
to get
\begin{equation*}
\begin{split}
\parallel \phi(x,.)&\parallel_{L^2(S^2)}^2\leq\\
&2\left(\frac{\pi}{2}-x\right)\sum_{(l,n)\in I}
\left(\int_0^{l\left(\frac{\pi}{2}-x\right)}\mid
  f'(t)\mid^{\frac{2}{1- 2m}}dt\right)^{1- 2m}l^{1+ 2m}\left(\mid
  u^l_{-,n}(\psi_-)\mid^2+
\mid u^l_{+,n}(\chi_-)\mid^2\right)\\
+
&2\left(\frac{\pi}{2}-x\right)\sum_{(l,n)\in I}
\left(\int_0^{l\left(\frac{\pi}{2}-x\right)}\mid
  f'(t)\mid^{\frac{2}{1+ 2m}}dt\right)^{1+ 2m}l^{1- 2m}\left(\mid
  u^l_{-,n}(\psi_+)\mid^2+
\mid u^l_{+,n}(\chi_+)\mid^2\right).
\end{split}
\end{equation*}
The dominated convergence theorem assures that $\phi$ satisfies
(\ref{faq}), and so $\Phi\in {\mathcal L}^2$. To achieve the proof, we
have to show that $H_m\Phi\in  {\mathcal L}^2$. We calculate :
\begin{equation*}
\begin{split}
H_m\Phi(x)=&
m\left(\frac{\pi}{2}-x\right)^{-m-1}\left(1-\frac{\frac{\pi}{2}-x}{\cos
    x}\right)
\sum_{(l,n)\in I}
f\left(l\left(\frac{\pi}{2}-x\right)\right)
\left(
\begin{array}{c}
u_{-,n}^l(\psi_-)T^l_{-\frac{1}{2},n}\\
u_{+,n}^l(\chi_-)T^l_{+\frac{1}{2},n}\\
-iu_{-,n}^l(\psi_-)T^l_{-\frac{1}{2},n}\\
iu_{+,n}^l(\chi_-)T^l_{+\frac{1}{2},n}
\end{array}
\right)\\
-&\left(\frac{\pi}{2}-x\right)^{-m}
\sum_{(l,n)\in I}
f'\left(l\left(\frac{\pi}{2}-x\right)\right)l
\left(
\begin{array}{c}
u_{-,n}^l(\psi_-)T^l_{-\frac{1}{2},n}\\
u_{+,n}^l(\chi_-)T^l_{+\frac{1}{2},n}\\
-iu_{-,n}^l(\psi_-)T^l_{-\frac{1}{2},n}\\
iu_{+,n}^l(\chi_-)T^l_{+\frac{1}{2},n}
\end{array}
\right)\\
+&\left(\frac{\pi}{2}-x\right)^{-m}\frac{1}{\sin x}
\sum_{(l,n)\in I}
f\left(l\left(\frac{\pi}{2}-x\right)\right)\left(l+\frac{1}{2}\right)
\left(
\begin{array}{c}
iu_{+,n}^l(\chi_-)T^l_{-\frac{1}{2},n}\\
-iu_{-,n}^l(\psi_-)T^l_{+\frac{1}{2},n}\\
u_{+,n}^l(\chi_-)T^l_{-\frac{1}{2},n}\\
u_{-,n}^l(\psi_-)T^l_{+\frac{1}{2},n}
\end{array}
\right)\\
-
&
m\left(\frac{\pi}{2}-x\right)^{m-1}\left(1-\frac{\frac{\pi}{2}-x}{\cos
    x}\right)
\sum_{(l,n)\in I}
f\left(l\left(\frac{\pi}{2}-x\right)\right)
\left(
\begin{array}{c}
u_{-,n}^l(\psi_+)T^l_{-\frac{1}{2},n}\\
u_{+,n}^l(\chi_+)T^l_{+\frac{1}{2},n}\\
iu_{-,n}^l(\psi_+)T^l_{-\frac{1}{2},n}\\
-iu_{+,n}^l(\chi_+)T^l_{+\frac{1}{2},n}
\end{array}
\right)\\
-&\left(\frac{\pi}{2}-x\right)^{m}
\sum_{(l,n)\in I}
f'\left(l\left(\frac{\pi}{2}-x\right)\right)l
\left(
\begin{array}{c}
u_{-,n}^l(\psi_+)T^l_{-\frac{1}{2},n}\\
u_{+,n}^l(\chi_+)T^l_{+\frac{1}{2},n}\\
iu_{-,n}^l(\psi_+)T^l_{-\frac{1}{2},n}\\
-iu_{+,n}^l(\chi_+)T^l_{+\frac{1}{2},n}
\end{array}
\right)\\
+&\left(\frac{\pi}{2}-x\right)^{m}\frac{1}{\sin x}
\sum_{(l,n)\in I}
f\left(l\left(\frac{\pi}{2}-x\right)\right)\left(l+\frac{1}{2}\right)
\left(
\begin{array}{c}
-iu_{+,n}^l(\chi_+)T^l_{-\frac{1}{2},n}\\
iu_{-,n}^l(\psi_+)T^l_{+\frac{1}{2},n}\\
u_{+,n}^l(\chi_+)T^l_{-\frac{1}{2},n}\\
u_{-,n}^l(\psi_+)T^l_{+\frac{1}{2},n}
\end{array}
\right).
\end{split}
\end{equation*}
In this sum, the leading terms have the form
$$
\Xi^{\pm}(x,\theta,\varphi)=\left(\frac{\pi}{2}-x\right)^{\pm m}\sum_{(l,n)\in I}
h\left(l\left(\frac{\pi}{2}-x\right)\right)\left(l+\frac{1}{2}\right) g_{l,n}^{\pm}(\theta,\varphi),
$$
where $h\in C^{\infty}_0([0,1[)$ and 
$$
\sum_{(l,n)\in I}\left(l+\frac{1}{2}\right)^{1\mp 2m}\parallel g_{l,n}^{\pm}\parallel^2_{L^2(S^2)}<\infty,\;\;\int_{S^2}g_{l,n}^{\pm}(\omega)\overline{g_{l',n'}^{\pm}(\omega)}d\omega=\delta_{l,l'}\delta_{n,n'}.
$$
Taking account of the support of $h$, we evaluate
\begin{equation*}
\begin{split}
\parallel \Xi^{\pm}\parallel_{L^2(]0,\pi[\times S^2)}^2&=
\sum_{(l,n)\in I}\left(l+\frac{1}{2}\right)^2\parallel
g_{l,n}^{\pm}\parallel_{L^2(S^2)}^2
\int_{\frac{\pi}{2}-\frac{1}{l}}^{\frac{\pi}{2}}\left(\frac{\pi}{2}-x\right)^{\pm
  2m}\left\vert
  h\left(l\left(\frac{1}{2}-x\right)\right)\right\vert^2dx\\
&\leq \int_0^1t^{\pm 2m}\mid h(t)\mid^2dt \sum_{(l,n)\in
  I}\left(l+\frac{1}{2}\right)^{1\mp 2m}\parallel
g_{l,n}^{\pm}\parallel_{L^2(S^2)}^2<\infty.
\end {split}
\end{equation*}

\fin
%%%%%%%%%%%%%%%%%%%%%%%%%%%%%%%%%%%%%%%%%%%%%%%%%%%%%%%%%%%%%%%%%%%%%%%%%%%%%%%%%%%%%%

{\it Proof of Theorem \ref{behcar}.}
Since the map $\mathbf S$ given by (\ref{S!}) satisfies (\ref{intert}),
 (\ref{regfip}) and (\ref{estcar}) follow from Proposition \ref{WSHS} and
(\ref{regfi}), and (\ref{estow}). Moreover, since
$\frac{\pi}{2}-x\sim\frac{1}{2}(1-\varrho)$, (\ref{deko}) and
(\ref{deku}) imply (\ref{dekop}) and
(\ref{dekup}). Now, if we put
$$
\Psi_{\pm}(\omega)=S(\theta,\varphi)\left(
\begin{array}{c}
\psi_{\pm}(\theta,\varphi)\\
\chi_{\pm}(\theta,\varphi)\\
\pm i\psi_{\pm}(\theta,\varphi)\\
\mp i \chi_{\pm}(\theta,\varphi)
\end{array}
\right),
$$
(\ref{dqfp}) and (\ref{polar}) are consequences respectively of
(\ref{dqf}), and :
$$
\left(\tilde\gamma^1\mp i
  Id\right)\Psi_{\pm}(\omega)=S(\theta,\varphi)\left(\gamma^1\mp iId\right)\left(
\begin{array}{c}
\psi_{\pm}(\theta,\varphi)\\
\chi_{\pm}(\theta,\varphi)\\
\pm i\psi_{\pm}(\theta,\varphi)\\
\mp i \chi_{\pm}(\theta,\varphi)
\end{array}
\right)=0.
$$
Finally, for any  $\Psi_-\in \left[H^{\frac{1}{2}+m}(S^2)\right]^4$,
$\Psi_{+}\in \left[H^{\frac{1}{2}-m}(S^2)\right]^4$ we define
$\Phi_{\pm}(\theta,\varphi):=S^*(\theta,\varphi)\Psi(\omega)\in{\mathcal
  W}^{\frac{1}{2}\mp m}$.  When $\Psi_{\pm}$ satisfy  (\ref{polar}),
then $\Phi_{\pm}$ have the form
$$
\Phi_{\pm}(\theta,\varphi)=\left(
\begin{array}{c}
\psi_{\pm}(\theta,\varphi)\\
\chi_{\pm}(\theta,\varphi)\\
\pm i\psi_{\pm}(\theta,\varphi)\\
\mp i \chi_{\pm}(\theta,\varphi)
\end{array}
\right),\;\;\psi_-\in W_-^{\frac{1}{2}+m},\;\;
\chi_-\in W_+^{\frac{1}{2}+m},\;\;\psi_{+}\in W_-^{\frac{1}{2}-m},\;\;\chi_{+}\in W_+^{\frac{1}{2}-m},
$$
and there exists $\Phi \in D(H_m)$
satisfying (\ref{dqf}) and (\ref{faq}). We conclude that
$\Psi:=\mathbf{S}\Phi$,
belongs to $D(\mathbf H_M)$ and satisfies (\ref{dqfp}) and
(\ref{faqp}). At last, Remark \ref{rembp} directly follows from
(\ref{regmo}), Proposition \ref{WSHS} and Remark \ref{remb}.

\fin
%%%%%%%%%%%%%%%%%%%%%%%%%%%%%%%%%%%%%%%%%%%%%%%%%%%%%%%%%%%%%%%%%%%%%%%%%%%%%%%%%%%%%%
We end this part by an important result of compactness :

\begin{Proposition}
Let $K$ be the set
\begin{equation}
K:=\left\{\Phi\in D(H_m),\;\;\parallel \Phi\parallel^2_{\mathcal
    L^2}+\parallel \Phi\parallel^2_{\mathcal L^2}\leq 1\right\} ;
  \label{cocop}
\end{equation}
Then, when $m>0$, $K$ is a compact of $\mathcal L^2$.
  \label{compak}
\end{Proposition}

{\it Proof of Proposition \ref{compak}.} We consider a sequence
$\left(\Phi^{\nu}\right)_{\nu\in\NN}$ in $K$. We write

$$
\Phi^{\nu}=\sum_{(l,n)\in I}
\left(
\begin{array}{c}
u^{l,\nu}_{1,n}T^l_{-\frac{1}{2},n}\\
u^{l,\nu}_{2,n}T^l_{+\frac{1}{2},n}\\
u^{l,\nu}_{3,n}T^l_{-\frac{1}{2},n}\\
u^{l,\nu}_{4,n}T^l_{+\frac{1}{2},n}
\end{array}
\right),
\;\;\;
H_m\Phi^{\nu}=\sum_{(l,n)\in I}
\left(
\begin{array}{c}
f^{l,\nu}_{1,n}T^l_{-\frac{1}{2},n}\\
f^{l,\nu}_{2,n}T^l_{+\frac{1}{2},n}\\
f^{l,\nu}_{3,n}T^l_{-\frac{1}{2},n}\\
f^{l,\nu}_{4,n}T^l_{+\frac{1}{2},n}
\end{array}
\right),
$$
and we have :
$$
\sum_{j=1}^4\sum_{(l,n)\in
  I}\parallel
u^{l,\nu}_{j,n}\parallel_{L^2(0,\frac{\pi}{2})}^2+\parallel
f^{l,\nu}_{j,n}\parallel_{L^2(0,\frac{\pi}{2})}^2\leq 1.
$$
The Banach-Alaoglu theorem assures that there exists $\Phi\in K$ and
a sub-sequence denoted $\left(\Phi^{\nu}\right)_{\nu\in\NN}$ again,
such that
$$
\Phi^{\nu}\rightharpoonup\Phi=\sum_{(l,n)\in I}
\left(
\begin{array}{c}
u^{l}_{1,n}T^l_{-\frac{1}{2},n}\\
u^{l}_{2,n}T^l_{+\frac{1}{2},n}\\
u^{l}_{3,n}T^l_{-\frac{1}{2},n}\\
u^{l}_{4,n}T^l_{+\frac{1}{2},n}
\end{array}
\right),
\;\;\;
H_m\Phi^{\nu}\rightharpoonup H_m\Phi=\sum_{(l,n)\in I}
\left(
\begin{array}{c}
f^{l}_{1,n}T^l_{-\frac{1}{2},n}\\
f^{l}_{2,n}T^l_{+\frac{1}{2},n}\\
f^{l}_{3,n}T^l_{-\frac{1}{2},n}\\
f^{l}_{4,n}T^l_{+\frac{1}{2},n}
\end{array}
\right)\;\;in\;\;\mathcal L^2-*,\;\;\nu\rightarrow\infty.
$$
Since for any $(l,n)\in I$, $j=1,..4$, $u^{l,\nu}_{j,n}\rightharpoonup
u^l_{j,n}$, $f^{l,\nu}_{j,n}\rightharpoonup
f^l_{j,n}$, in $L^2(0,\frac{\pi}{2})-*$ as $\nu\rightarrow\infty$, 
we deduce from (\ref{eqvpl}), (\ref{equvpl}), (\ref{eqppl}) and
(\ref{equpl}), that
$$
\forall x\in[0, \frac{\pi}{2}],\;\;u^{l,\nu}_{j,n}(x)\rightarrow
u^l_{j,n}(x),\;\;\sup_{\nu}\sup_{x\in[0,\frac{\pi}{2}]}\mid u^{l,\nu}_{j,n}(x)\mid<\infty.
$$
Therefore
\begin{equation}
\parallel
u^{l,\nu}_{j,n}-u^l_{j,n}\parallel_{L^2(0,\frac{\pi}{2}]}\rightarrow 0,\;\;\nu\rightarrow\infty.
  \label{lio}
\end{equation}
Moreove, since $m>0$, (\ref{estow}) implies :
\begin{equation}
\sup_{\nu}\sum_{(l,n)\in I}\left(l+\frac{1}{2}\right)^2\sum_{j=1}^4\parallel u_{j,n}^{l,\nu}-u^l_{j,n}\parallel_{L^2(0,\frac{\pi}{2})}^2<\infty.
  \label{supi}
\end{equation}
For $l\in\NN+\frac{1}{2}$ we put
$$
\varepsilon^{l,\nu}:=\sum_{j=1}^4\sum_{n=-l}^l\parallel
u_{j,n}^{l,\nu}-u^l_{j,n}\parallel_{L^2(0,\frac{\pi}{2})}^2.
$$
(\ref{lio}) and (\ref{supi}) show that
$$
\forall l\in\NN+\frac{1}{2},\;\;\varepsilon^{l,\nu}\rightarrow 0,\;\;\nu\rightarrow\infty,\;\;A:=\sup_{\nu}\sum_{l\in\NN+\frac{1}{2}}\left(l+\frac{1}{2}\right)^2\varepsilon^{l,\nu}<\infty.
$$
Since $\varepsilon^{l,\nu}\leq A\left(l+\frac{1}{2}\right)^{-2}$,
the dominated convergence theorem implies that
$\sum_l\varepsilon^{l,\nu}\rightarrow 0$, as $\nu\rightarrow\infty$,
that is to say, $\Phi^{\nu}$ strongly tends to $\Phi$ in $\mathcal L^2$.

\fin

%%%%%%%%%%%%%%%%%%%%%%%%%%%%%%%%%%%%%%%%%%%%%%%%%%%%%%%%%%%%%%%%%%%%%%%%%%%%%%%%%%%%%%%%%%%%%%%%%%%%%%%%%%%%%%%%%%%%%%%%%%%%%%%%%%%%%%%%%%%%%%%%%%%%%%%%%%%%%%%%%

\section{Self-adjoint extensions}

%%%%%%%%%%%%%%%%%%%%%%%%%%%%%%%%%%%%%%%%%%%%%%%%%%%%%%%%%%%%%%%%%%%%%%%%%%%%%%%%%%%%
When $0<m<\frac{1}{2}$, we define the
linear map
\begin{equation*}
\Gamma:\;\Phi\in D(H_m)\longmapsto \Gamma(\Phi)=
\left(
\begin{array}{c}
\psi_-\\
\chi_-\\
\psi_+\\
\chi_+
\end{array}
\right)\in{\mathcal W}^{\frac{1}{2}},
\end{equation*}
where $\psi_{\pm}$ and $\chi_{\pm}$ are given by (\ref{dqf}), and we
put $\Gamma(\Phi)=0$ when $\frac{1}{2}\leq m$. We note that Theorem
\ref{proreg} assures that
\begin{equation}
\forall m\in]0,\frac{1}{2}[,\;\;W^{\frac{1}{2}+m}_-\times
W^{\frac{1}{2}+m}_+\times W^{\frac{1}{2}-m}_-\times
W^{\frac{1}{2}-m}_+\subset \Gamma\left(D(H_m)\right).
  \label{surj}
\end{equation}
We introduce the matrix
\begin{equation*}
Q:=-\gamma^0\gamma^5=
\left(
\begin{array}{cccc}
0&0&-1&0\\
0&0&0&-1\\
1&0&0&0\\
0&1&0&0
\end{array}
\right).
\end{equation*}

The basic tool is a nice Green formula : 
\begin{Lemma}
Given $0<m$, for any $\Phi,\tilde\Phi\in D(H_m)$ we have
\begin{equation}
<H_m\Phi,\tilde\Phi>_{\mathcal L^2}-<\Phi,H_m\tilde\Phi>_{\mathcal
  L^2}=2<\Gamma(\Phi),Q\Gamma(\tilde\Phi)>_{\mathcal W^0}.
  \label{green}
\end{equation}
  \label{corg}
\end{Lemma}

{\it Proof of Lemma \ref{corg}.} (\ref{estow}) assures that any
$\Phi\in D(H_m)$ belongs to $L^2\left([0,\frac{\pi}{2}]_x;{\mathcal
    W}^1\right)$, hence for any $\varepsilon>0$, $\Phi\in
H^1\left(]\varepsilon,\frac{\pi}{2}-\varepsilon[_x;{\mathcal
    W}^0\right)$. Since $(\mathbf D, \mathcal W^1)$ is selfadjoint on
$\mathcal W^0$, we evaluate
$$
<H_m\Phi,\tilde\Phi>_{\mathcal L^2}-<\Phi,H_m\tilde\Phi>_{\mathcal
  L^2}=
\lim_{\varepsilon\rightarrow
  0}<i\gamma^0\gamma^1\Phi(\frac{\pi}{2}-\varepsilon),\tilde\Phi(\frac{\pi}{2}-\varepsilon)>_{\mathcal W^0}-<i\gamma^0\gamma^1\Phi(\varepsilon),\tilde\Phi(\varepsilon)>_{\mathcal W^0},
$$
and taking account of  (\ref{compo}),  (\ref{deko}), (\ref{deku}),
(\ref{dqf}) and (\ref{faq}) we get (\ref{green}).
\fin

We now investigate the self-adjoint extensions $(\mathcal H, D(\mathcal
H))$ of $H_m$, with $C^{\infty}_0(]0,\frac{\pi}{2}[_x\times
]0,\pi[_{\theta}\times ]0,2\pi[_{\varphi};\CC^4)\subset D(\mathcal
H)$. The adjoint $\mathcal
H^*$ is just $H_m$ with domain $D(\mathcal H^*)\subset D(H_m)$, and we
have :
$$
\forall\Phi\in D(\mathcal H),\;\;\forall \tilde\Phi\in D(\mathcal H^*),\;
\;<\Gamma(\Phi),Q\Gamma(\tilde\Phi)>_{\mathcal W^0}=0.
$$

When $m\geq\frac{1}{2}$ we immediately obtain a first result of self-adjointness of $H_m$ on
$\mathcal L^2$ :
%%%%%%%%%%%%%%%%%%%%%%%%%%%%%%%%%%%%%%%%%%%%%%%%%%%%%%%%%%%%%%%%%%%%%%%%%%%%%%%%%%%%%

\begin{Proposition}
When $\frac{1}{2}\leq m$, $H_m$ is essentially self-adjoint on
$\left[C^{\infty}_0\left(]0,\frac{\pi}{2}[_x\times]0,\pi[_{\theta}\times]0,2\pi[_{\varphi}\right)\right]^4$.
  \label{sefo}
\end{Proposition}

%%%%%%%%%%%%%%%%%%%%%%%%%%%%%%%%%%%%%%%%%%%%%%%%%%%%%%%%%%%%%%%%%%%%%%%%%%%%

{\it Proof of Proposition \ref{sefo}.} Let $\mathcal H$ be the operator
defined by the differential operator $H_m$ endowed with the domain
$D(\mathcal H)=C^{\infty}_0(]0,\frac{\pi}{2}[_x\times
]0,\pi[_{\theta}\times ]0,2\pi[_{\varphi};\CC^4)$. On the one hand,
$\mathcal H$ is obviously symmetric, and on the other hand, its adjoint $\mathcal
H^*$ is just $H_m$ with domain $D(H_m)$. Any $\Phi_{\pm}$ be in
$D(H_m)$ such that
${\mathcal H}^*\Phi_{\pm}\pm i\Phi_{\pm}=0$, satisfies
$$
\mp2i\parallel \Phi\parallel^2_{\mathcal L^2}=
<H_m\Phi_{\pm},\Phi_{\pm}>_{\mathcal L^2}-<\Phi_{\pm},H_m\Phi_{\pm}>_{\mathcal
  L^2},
$$
and we conclude by (\ref{green}) that $\Phi_{\pm}=0$.

\fin

%%%%%%%%%%%%%%%%%%%%%%%%%%%%%%%%%%%%%%%%%%%%%%%%%

When $0<m<\frac{1}{2}$, the situation is much more interesting : there
exists a lot of self-adjoint realizations of $H_m$. First, we introduce
the operators $\mathcal H_{MIT}$ and ${\mathcal H}_{CHI}$ respectively associated
with the {\it MIT-bag} and the {\it Chiral} boundary conditions. They are defined as $H_m$ endowed with the
domains
\begin{equation}
D\left(\mathcal H_{MIT}\right):=\left\{\Phi\in D(H_m);\;\;\parallel
  \gamma^1\Phi(x,.)+i\Phi(x,.)\parallel_{\mathcal W^0}=o\left(\sqrt{\frac{\pi}{2}-x}\right),\;\;x\rightarrow\frac{\pi}{2}\right\},
  \label{mit}
\end{equation}
\begin{equation}
D\left(\mathcal H_{CHI}\right):=\left\{\Phi\in D(H_m);\;\;\parallel
  \gamma^1\Phi(x,.)-i\Phi(x,.)\parallel_{\mathcal W^0}=o\left(\sqrt{\frac{\pi}{2}-x}\right),\;\;x\rightarrow\frac{\pi}{2}\right\}.
  \label{mitchi}
\end{equation}
In fact these asymptotic conditions are reduced to linear constraints
on the asymptotic profiles $\Phi_{\pm}$ :
 We check by (\ref{dqf}) that
$$
\gamma^1\Phi(x,\theta,\varphi)\pm i\Phi(x,\theta,\varphi)=\pm 2i\left(\frac{\pi}{2}-x\right)^{\pm m}
\left(
\begin{array}{c}
\psi_{\pm}(\theta,\varphi)\\
\chi_{\pm}(\theta,\varphi)\\
\pm i\psi_{\pm}(\theta,\varphi)\\
\mp i\chi_{\pm}(\theta,\varphi)
\end{array}
\right)
+\left(\gamma^1\pm i\right)\varphi(x,\theta,\varphi).
$$
Thus (\ref{faq}) implies that
$$
D\left(\mathcal H_{MIT}\right)=\left\{\Phi\in D(H_m);\;\;\psi_+=\chi_+=0\right\},
 $$
$$
D\left(\mathcal H_{CHI}\right)=\left\{\Phi\in D(H_m);\;\;\psi_-=\chi_-=0\right\}.
$$

%%%%%%%%%%%%%%%%%%%%%%%%%%%%%%%%%%%%%%%%%%%%%%%%%%%%%%%%%%%%%%%%%%%%%%%%%%%%%%%%%%%%%%%%%%%%%%

We now construct a large family of self-adjoint extensions that are
non-local generalizations of the {\it MIT-bag} and {\it Chiral} conditions. We consider
 densely defined self-adjoint operators $(A^{\pm},D(A^{\pm}))$ on $L^2(S^2)\times L^2(S^2)$,
satisfying
\begin{equation}
W^{\frac{1}{2}}_-\times W^{\frac{1}{2}}_+\subset
D(A^{-}),\;\;D(A^+)=L^2(S^2)\times L^2(S^2),
  \label{dapmm}
\end{equation}
\begin{equation}
A^{\pm}\left(C^{\infty}_0(]0,\pi[\times]0,2\pi[;\CC^2\right)\subset
W^{\frac{1}{2}\pm m}_-\times W^{\frac{1}{2}\pm m}_+.
  \label{hypa}
\end{equation}
We  introduce
the operators $\mathcal H_{A^{\pm}}$ defined as $H_m$ endowed with the
domain
\begin{equation*}
D\left(\mathcal H_{A^{\pm}}\right):=\left\{\Phi\in D(H_m);\;\;
\left(
\begin{array}{c}
\psi_{\mp}\\
\chi_{\mp}
\end{array}
\right)
=
A^{\pm}
\left(\begin{array}{c}
\psi_{\pm}\\
\chi_{\pm}
\end{array}
\right)
\right\}.
\end{equation*}
In particular, we have $\mathcal H_{A^-=0}=\mathcal H_{MIT}$ and
$\mathcal H_{A^+=0}=\mathcal H_{CHI}$.

\begin{Proposition}
When $0<m<\frac{1}{2}$, $\mathcal H_{A^+}$ and $\mathcal H_{A^-}$ are self-adjoint on
$\mathcal L^2$.
  \label{mAmA}
\end{Proposition}

{\it Proof of Proposition \ref{mAmA}.} Since $A^{\pm}$ are
self-adjoint, we have for $\Phi$, $\tilde\Phi\in D(\mathcal
H_{A^{\pm}})$ :
\begin{equation}
<\Gamma(\Phi),Q\Gamma(\tilde\Phi)>_{\mathcal W^0}=
\left<
\left(
\begin{array}{c}
\psi_{\pm}\\
\chi_{\pm}
\end{array}
\right),
\left(
\begin{array}{c}
\tilde\psi_{\mp}\\
\tilde\chi_{\mp}
\end{array}
\right)
-A^{\pm}
\left(
\begin{array}{c}
\tilde\psi_{\pm}\\
\tilde\chi_{\pm}
\end{array}
\right)
\right>_{L^2(S^2;\CC^2)}=0.
  \label{fpp}
\end{equation}
Therefore $\mathcal H_{A^{\pm}}$ are symmetric. Moreover, given $\tilde\Phi\in D(\mathcal
H_{A^{\pm}}^*)$, we have (\ref{fpp})  for any $\Phi\in D(\mathcal
H_{A^{\pm}})$ again. For all
$\psi_{\pm}$, $\chi_{\pm}\in C^{\infty}_0(]0,\pi[\times ]0,2\pi[)$, 
 (\ref{surj}) and (\ref{hypa}) assure there exists $\Phi\in D(\mathcal
H_{A^{\pm}})$ such that
$$
\Gamma(\Phi)=\left(A^+(\psi_+,\chi_+),\psi_+,\chi_+\right)\;\;or\;\;\left(\psi_-,\chi_-,A^-(\psi_-,\chi_-)\right).
$$
Therefore 
$$
\left<
\left(
\begin{array}{c}
\psi_{\pm}\\
\chi_{\pm}
\end{array}
\right),
\left(
\begin{array}{c}
\tilde\psi_{\mp}\\
\tilde\chi_{\mp}
\end{array}
\right)
-A^{\pm}
\left(
\begin{array}{c}
\tilde\psi_{\pm}\\
\tilde\chi_{\pm}
\end{array}
\right)
\right>_{L^2(S^2;\CC^2)}=0.
$$
We conclude that  $\tilde\Phi\in D(\mathcal
H_{A^{\pm}})$.

\fin

Finally we consider the operators $\mathcal H_{APS}$, $\mathcal
H_{mAPS}$ associated with the {\it APS} and {\it mAPS}
boundary conditions :

$$
D\left(\mathcal H_{APS}\right):=\left\{\Phi\in
  D(H_m);\;\;\parallel\mathbf K_+\Phi(x,.)\parallel_{\mathcal W^0}=o\left(\sqrt{\frac{\pi}{2}-x}\right),\;\;x\rightarrow\frac{\pi}{2}\right\},
$$
$$
D\left(\mathcal H_{mAPS}\right):=\left\{\Phi\in
  D(H_m);\;\;\parallel\mathbf K_+\left(Id+\gamma^1\right)\Phi(x,.)\parallel_{\mathcal W^0}=o\left(\sqrt{\frac{\pi}{2}-x}\right),\;\;x\rightarrow\frac{\pi}{2}\right\},
$$
where $\mathbf K_+$ is defined by (\ref{kaps}).

\begin{Proposition}
When $0<m<\frac{1}{2}$, we have
$$
D\left(\mathcal H_{APS}\right)=D\left(\mathcal H_{mAPS}\right)=\left\{\Phi\in
  D(H_m);\;\;\mathbf K_+\Phi_+=\mathbf K_+\Phi_-=0\right\},
 $$
and $\mathcal H_{APS}=\mathcal H_{mAPS}$  is self-adjoint on
$\mathcal L^2$.
  \label{KKK}
\end{Proposition}

{\it Proof of Proposition \ref{KKK}.} By (\ref{dqf}),
we have
\begin{equation*}
\begin{split}
\mathbf K_+&\Phi(x)\\
=&\left(\frac{\pi}{2}-x\right)^{-m}
\left(
\begin{array}{c}
1\\
-i\mathcal J^*\\
-i\\
\mathcal J^*
\end{array}
\right)
(\psi_-+i\mathcal J \chi_-)
+
\left(\frac{\pi}{2}-x\right)^{m}
\left(
\begin{array}{c}
1\\
i\mathcal J^*\\
i\\
\mathcal J^*
\end{array}
\right)
(\psi_+-i\mathcal J \chi_+)
+\mathbf K_+\varphi(x),
\end{split}
\end{equation*}

\begin{equation*}
\begin{split}
\mathbf K_+&\left(Id+\gamma^1\right)\Phi(x)\\
=&(1-i)\left(\frac{\pi}{2}-x\right)^{-m}
\left(
\begin{array}{c}
1\\
-i\mathcal J^*\\
-i\\
\mathcal J^*
\end{array}
\right)
(\psi_-+i\mathcal J \chi_-)
+
(1+i)\left(\frac{\pi}{2}-x\right)^{m}
\left(
\begin{array}{c}
1\\
i\mathcal J^*\\
i\\
\mathcal J^*
\end{array}
\right)
(\psi_+-i\mathcal J \chi_+)\\
&+\mathbf K_+\left(Id+\gamma^1\right)\varphi(x),
\end{split}
\end{equation*}
thus we deduce from (\ref{faq}) that
\begin{equation}
\begin{split}
\parallel\mathbf
 K_+\Phi(x,.)\parallel_{\mathcal
  W^0}=o\left(\sqrt{\frac{\pi}{2}-x}\right)
\Leftrightarrow&
\parallel\mathbf
K_+\left(Id+\gamma^1\right)\Phi(x,.)\parallel_{\mathcal
  W^0}=o\left(\sqrt{\frac{\pi}{2}-x}\right)\\
\Leftrightarrow &\psi_{\pm}=\pm i\mathcal J \chi_{\pm}\\
\Leftrightarrow &\mathbf K_+\Phi_+=\mathbf K_+\Phi_-=0.
\end{split}
  \label{apsmp}
\end{equation}
This equality assures that for $\Phi,\tilde\Phi\in D\left(\mathcal
  H_{APS}\right)$, we have $<\Gamma\Phi,Q\Gamma\Phi>_{\mathcal
  W^0}=0$, i.e. $\mathcal H_{APS}$ is symmetric. Moreover, for any $\Phi\in D\left(\mathcal
  H_{APS}\right)$,  $\tilde\Phi\in D\left(\mathcal
  H_{APS}^*\right)$, we have
\begin{equation}
<\chi_+,\tilde \chi_--i\mathcal J^*\tilde\psi_->_{L^2(S^2)}-
<\chi_-,\tilde \chi_++i\mathcal J^*\tilde\psi_+>_{L^2(S^2)}=0.
  \label{kikio}
\end{equation}
Since $C^{\infty}_0(]0,\pi[\times]0,2\pi[)\subset W_{+}^1$, for any
$\chi_{\pm}\in C^{\infty}_0(]0,\pi[\times]0,2\pi[)$, $\mathcal
J\chi_{\pm}$ belongs to $W^1_-$ and by (\ref{surj}) there exists
$\Phi\in D(H_m)$ such that
$$
\Gamma(\Phi)= 
\left(
\begin{array}{c}
-i\mathcal J \chi_-\\
\chi_-\\
 i\mathcal J \chi_+\\
 \chi_+
\end{array}
\right).
$$
But such a $\Phi$ satisfies (\ref{apsmp}), that means that $\Phi$ is
in the domain of $\mathcal H_{APS}$. Since $\chi_{\pm}$ are arbitrary,
(\ref{kikio}) implies
$$
\tilde \chi_{\pm}\pm i\mathcal J^*\tilde\psi_{\pm}=0,
$$
that is equivalent to (\ref{apsmp}). We conclude that $\tilde\Phi\in
D\left(\mathcal H_{APS}\right)$.
\fin

The remainder of the article is devoted to the demonstrations of
theorems of Part 3. As we have explained above, it is sufficient to
consider only the case $M>0$, since the chiral transform chiral changes the sign of the mass.\\
%%%%%%%%%%%%%%%%%%%%%%%%%%%%%%%%%%%%%%%%%%%%%%%%%%%%%%%%%%%%%%%%%%%%%%%%%%%%%%%%%%%%%%

{\it Proof of Theorem \ref{maintheo}.} We denote $\mathbb
  H$ the operator $\mathbf H_M$ endowed with the domain $D(\mathbb H):=\left[C^{\infty}_0(\BB)\right]^4$.
Since $\mathbf H_M=\mathbf S H_m\mathbf S^{-1}$, Proposition \ref{sefo}
assures that  $\mathbf H_M$ is essentially self-adjoint on $\mathbf
S\left(C^{\infty}_0\left(\left]0,\frac{\pi}{2}\right[_x\times
      ]0,\pi[_{\theta}\times
      ]0,2\pi[_{\varphi};\CC^4\right)\right)$ when $ M\geq
  \sqrt{\frac{\Lambda}{12}}$. Proposition \ref{WSHS} and the Sobolev
  imbedding Theorem imply that this set is
  included in  $\left[C^{\infty}_0(\BB)\right]^4$. Since $\mathbb
  H$ is symmetric, we deduce that it is essentially self-adjoint.\\

To determine its domain and establish the elliptic estimate, we prove
an inequality of Hardy type. Given a real valued function $f\in
C^1_0(]0,1[)$, an integration by part gives :
\begin{equation*}
\begin{split}
\int_0^1f^2(\varrho)\frac{\varrho^2}{(1-\varrho^2)^2}d\varrho=&-\frac{1}{2}\int_0^1f^2(\varrho)\frac{\varrho}{1-\varrho^2}d\varrho+\int_0^1\left(f(\varrho)\frac{\varrho}{1-\varrho^2}\right)\left(\varrho
  f'(\varrho)\right)d\varrho\\
\leq&\frac{1}{2}\int_0^1f^2(\varrho)\frac{\varrho^2}{(1-\varrho^2)^2}d\varrho
+\frac{1}{2}\int_0^1\varrho^2f'^2(\varrho)d\varrho,
\end{split}
\end{equation*}
hence by density we get that
$\frac{1}{1-\varrho}\Psi\in \mathbf L^2$ when $\Psi\in
\left[H^1_0(\BB)\right]^4$, and we have  the following Hardy estimate :
\begin{equation}
\forall \Psi\in H^1_0(\mathbb B),\;\;\int_{\mathbb B}\mid \Psi(\mathbf
x)\mid^2\frac{1}{(1-\mid\mathbf x\mid^2)^2}d\mathbf x\leq
\int_{\mathbb B}\mid\nabla_{\mathbf x}\Psi(\mathbf x)\mid^2d\mathbf x.
  \label{hardy}
\end{equation}
Thus we see that $\left[H^1_0(\BB)\right]^4\subset
D(\mathbf H_M)$ and the graph norm of $\mathbf H_M$ is bounded by the
$H^1_0$ norm. Conversely, for
$\Psi\in\left[C^{\infty}_0(\BB)\right]^4$, we use  the Fourier transform of $\Psi$, the Parseval formula and
the anticommutations relations (\ref{comanti}) to remark that
$$
\int_{\BB}\sum_{1\leq i<j\leq 3}\partial_i\Psi^*\gamma^i\gamma^j\partial_j\Psi+
\partial_j\Psi^*\gamma^j\gamma^i\partial_i\Psi \;d\mathbf x=0,
$$
then we calculate
\begin{equation*}
\begin{split}
\int_{\BB}\left\vert\gamma^j\partial_j\Psi+\frac{2iM}{1-\varrho^2}\sqrt{\frac{3}{\Lambda}}\Psi\right\vert^2&d\mathbf
x
=\int_{\BB}\mid\nabla_{\mathbf x}\Psi\mid^2+
\frac{12M^2}{\Lambda(1-\varrho^2)^2}\mid\Psi\mid^2
+\frac{4iM}{(1-\varrho^2)^2}\sqrt{\frac{3}{\Lambda}}x_j\Psi^*\gamma^j\Psi
\;d\mathbf x.
\end{split}
\end{equation*}
Therefore the Hardy inequality (\ref{hardy}) shows that when  $M>
\sqrt{\frac{\Lambda}{12}}$, elliptic estimate (\ref{ellip}) holds :
$$
\parallel \mathbf H_M\Psi\parallel_{\mathbf L^2}^2\geq\left(1-M\sqrt{\frac{12}{\Lambda}}\right)^2\int_{\BB}\mid\nabla_{\mathbf x}\Psi\mid^2d\mathbf x,
$$
and the $H^1_0$-norm on $\left[C^{\infty}_0(\BB)\right]^4$ is bounded by
the graph norm of $\mathbf H_M$. Since $\HH$ is essentially
self-adjoint, we have $\HH^*=\overline\HH$. On the one hand $D(\HH^*)=D(\mathbf
H_M)$. On the other hand $D(\overline\HH)$ is the closure of
$\left[C^{\infty}_0(\BB)\right]^4$ for the graph norm. We conclude
that $D(\mathbf
H_M)=\left[H^1_0(\BB)\right]^4$ when $M> \sqrt{\frac{\Lambda}{12}}$
and the first part of the
  Theorem is proved.

Now when $ 0<M< \sqrt{\frac{\Lambda}{12}}$, and
  $\mathbf A^{\pm}$ satisfy (\ref{dapm}) and (\ref{capm}), then
  $A^{\pm}=S_{11}^*\mathbf A^{\pm} S_{11}$ where $S_{11}$ is defined
  by (\ref{sunun}), satisfy (\ref{dapmm}) and (\ref{hypa}). We deduce
  fom Proposition \ref{mAmA} that $\mathbb H_{\mathbf A^{\pm}}=\mathbf
  S\mathcal H_{ A^{\pm}}\mathbf
  S^{-1}$ is self-adjoint.
On the other hand, we have $\mathbb H_{\mathcal B_{APS}}=\mathbf S\mathcal
H_{APS}\mathbf S^{-1}=\mathbf S H_{mAPS}\mathbf S^{-1}=\mathbb H_{\mathcal
  B_{mAPS}}$ that is self-adjoint by Proposition \ref{KKK}.

Finally $\left\{\Psi\in D(\mathbf H_M),\;\;\parallel\Psi\parallel_{\mathbf
      L^2}^2+\parallel\mathbf H_M\Psi\parallel_{\mathbf
      L^2}^2\leq 1\right\}$ is equal to $\mathbf S K$ where $K$
  defined by (\ref{cocop}) is compact by Proposition \ref{compak}. We
  conclude that the resolvent of any self-adjoint realization of
  $\mathbf H_M$ is compact.
\fin
%%%%%%%%%%%%%%%%%%%%%%%%%%%%%%%%%%%%%%%%%%%%%%%%%%%%%%%%%%%%%%%%%%%%%%%%%%%%%%%%%%%%%%

%%%%%%%%%%%%%%%%%%%%%%%%%%%%%%%%%%%%%%%%%%%%%%%%%%%%%%%%%%%%%%%%%%%%%%%%%%%%%%%%%%%%%%

{\it Proof of Theorem \ref{theoevol}.}
Theorem \ref{maintheo} provides a lot of solutions of the initial
value problem :  if $\mathbb H$ is a self-adjoint realization of
$\mathbf H_M$, $\Psi(t)=e^{it\sqrt{\frac{\Lambda}{3}}\mathbb H}\Psi_0$
is a solution of (\ref{eqcar}), (\ref{reg}), (\ref{CI}) and (\ref{conserve}).

Since the maximal globally
hyperbolic domain in  ${\mathcal
  E}$ including $\{t=0\}\times [0,\frac{\pi}{2}[_x\times S^2_{\theta,\varphi}$ is given by
$
0\leq \mid t\mid <\sqrt{\frac{3}{\Lambda}}\left(\frac{\pi}{2}-x\right),
$
the maximal globally
hyperbolic domain in  ${\mathcal
  M}$ including $\{t=0\}\times \RR^3$ is defined by the same relation,
that is in $(t,\mathbf x)$ coordinates :
$
0\leq 
\mid t\mid <\sqrt{\frac{3}{\Lambda}}\left(\frac{\pi}{2}-2\arctan\varrho\right)
 $. We show that all the solutions are equal in this domain. Given
 $\Psi$ satisfying (\ref{eqcar}), (\ref{reg}), we introduce for all
 $\varepsilon>0$
$$
\Psi_{\varepsilon}(t)=\frac{1}{\varepsilon}\int_t^{t+\varepsilon}\Psi(s)ds.
$$
It is clear that $\Psi_{\varepsilon}\in C^1(\RR_t;\mathbf L^2)$,
$\Psi_{\varepsilon}\rightarrow \Psi$ in $ C^0(\RR_t;\mathbf L^2)$ as
$\varepsilon\rightarrow 0$. Moreover we can see that $\Psi_{\varepsilon}$ is solution of
(\ref{eqcar}), thus
$\mathbf H_M\Psi_{\varepsilon} \in C^0(\RR_t;\mathbf L^2)$ and
$$
\frac{\partial}{\partial
  t}\left(\sqrt{\frac{3}{\Lambda}}\mid\Psi_{\varepsilon}\mid^2\right)+\sum_{j=1}^3\frac{1+\varrho^2}{2}\frac{\partial}{\partial x^j}\left(\Psi_{\varepsilon}^*\gamma^0\gamma^j\Psi_{\varepsilon}\right)=0.
$$
We integrate this equality on $\left\{(t,\mathbf x),\;0\leq t\leq
  T,\;\varrho\leq
  \tan\left(\frac{\pi}{4}-T\sqrt{\frac{\Lambda}{12}}\right)\right\}$
where $0<T<\frac{\pi}{2}\sqrt{\frac{3}{\Lambda}}$, and applying the
Green formula we get
\begin{equation*}
\begin{split}
\int_{\varrho\leq \tan(\frac{\pi}{4}-T\sqrt{\frac{\Lambda}{12}})}\mid
\Psi_{\varepsilon}(T,\mathbf x)\mid^2d\mathbf x=&\int_{\mathbb B}\mid
\Psi_{\varepsilon}(0,\mathbf x)\mid^2d\mathbf x\\
&-
\int_{0\leq t=\sqrt{\frac{3}{\Lambda}}(\frac{\pi}{2}-2\arctan\varrho)\leq
  T}\mid\Psi(t,\mathbf
x)\mid^2-\frac{x_j}{\varrho}\Psi^*\gamma^0\gamma^j\Psi(t,\mathbf x)d\sigma.
\end{split}
\end{equation*}
The last integral is non-negative since $\mid x_j\gamma^j\Psi\mid\leq
\varrho\mid\Psi\mid$, and taking the limit as $\varepsilon\rightarrow 0$, we
obtain
$$
\int_{\varrho\leq \tan(\frac{\pi}{4}-T\sqrt{\frac{\Lambda}{12}})}\mid
\Psi(T,\mathbf x)\mid^2d\mathbf x\leq\int_{\mathbb B}\mid
\Psi(0,\mathbf x)\mid^2d\mathbf x.
$$
We conclude that $\Psi=0$ for $
0\leq 
\mid t\mid <\sqrt{\frac{3}{\Lambda}}\left(\frac{\pi}{2}-2\arctan\varrho\right)
 $ if $\Psi_0=0$. Finally when  $ M\geq \sqrt{\frac{\Lambda}{12}}$, we
 use the fact that $\mathbf H_M$ is self-adjoint to write
$$
\frac{d}{dt}\left(e^{-it\sqrt{\frac{\Lambda}{3}}\mathbf H_M}\Psi_{\varepsilon}(t)\right)=e^{-it\sqrt{\frac{\Lambda}{3}}\mathbf H_M}\left(-i\sqrt{\frac{\Lambda}{3}}\mathbf H_M\Psi_{\varepsilon}(t)+\partial_t\Psi_{\varepsilon}(t)\right)=0,
$$
and we deduce that
$\Psi_{\varepsilon}(t)=e^{it\sqrt{\frac{\Lambda}{3}}\mathbf
  H_M}\Psi_{\varepsilon}(0)$. Taking the limit in $\varepsilon$ again,
we conclude that $\Psi(t)=e^{it\sqrt{\frac{\Lambda}{3}}\mathbf
  H_M}\Psi_0$.
\fin
%%%%%%%%%%%%%%%%%%%%%%%%%%%%%%%%%%%%%%%%%%%%%%%%%%%%%%%%%%%%%%%%%%%%%%%%%%%%%%%%%%%%%%
%%%%%%%%%%%%%%%%%%%%%%%%%%%%%%%%%%%%%%%%%%%%%%%%%%%%%%%%%%%%%%%%%%%%%%%%%%%%%%%%%%%%%%

{\it Proof of Theorem \ref{equipat}.}
Since the spectrum of $\mathbb H$ is dicrete, and $0$ is not an
eigenvalue when $M>0$, there exists  an orthonormal
basis of eigenvectors,
$\left(\Psi_k\right)_{k\in\NN}$, with $\mathbf H_M\Psi_k=\lambda_k\sqrt{\frac{\Lambda}{3}}\Psi_k$,
$\lambda_k\in\RR^*$. Now the crucial point is that
\begin{equation}
\int_{\mathbb B}\Psi_k^*\gamma^0\gamma^5\Psi_k(\mathbf x)d\mathbf x=0.
  \label{ortop}
\end{equation}
To see that, we note that $\mathbf
H_M\gamma^0\gamma^5=-\gamma^0\gamma^5\mathbf H_M$, and we write
\begin{equation*}
\begin{split}
<\Psi_k,\gamma^0\gamma^5\Psi_k>_{\mathbf
  L^2}&=\frac{1}{\lambda_k}<\mathbf
H_M\Psi_k,\gamma^0\gamma^5\Psi_k>_{\mathbf L^2}\\
&=-\frac{1}{\lambda_k}<\Psi_k,\gamma^0\gamma^5\mathbf
H_M\Psi_k>_{\mathbf L^2}\\
&=-<\Psi_k,\gamma^0\gamma^5\Psi_k>_{\mathbf L^2}.
\end{split}
\end{equation*}
We can expand $\Psi$ on
this basis :
$$
\Psi(t,\mathbf x)=\sum_{k\in\NN}
c_ke^{i\lambda_kt}\Psi_k(\mathbf
x),\;\;c_k\in\CC,\;\;\sum_{k\in\NN}\mid c_k\mid^2<\infty,
$$
and taking advantage of (\ref{ortop}) we evaluate
$$
\frac{1}{T}\int_0^T\int_{\mathbb
  B}\Psi^*\gamma^0\gamma^5\Psi(t,\mathbf x)d\mathbf
xdt=\sum_{\lambda_p\neq\lambda_q}c_pc_q^*\frac{e^{i(\lambda_p-\lambda_q)T}-1}{i(\lambda_p-\lambda_q)T}\int_{\mathbb
  B}\Psi_q^*\gamma^0\gamma^5\Psi_p(\mathbf x)d\mathbf x.
$$
The dominated convergence theorem assures that this sum tends to $0$
as $T\rightarrow\infty$.
\fin
%%%%%%%%%%%%%%%%%%%%%%%%%%%%%%%%%%%%%%%%%%%%%%%%%%%%%%%%%%%%%%%%%%%%%%%%%%%%%%%%%%%%
%%%%%%%%%%%%%%%%%%%%%%%%%%%%%%%%%%%%%%%%%%%%%%%%%%%%%%%%%%%%%%%%%%%%%%%%%%%%%%%%%%%%

\section{Appendix. Breitenlohner-Freedman bounds for the scalar
  waves}
We consider the Klein-Gordon equation on the Anti-de Sitter space-time
$$
\mid g\mid^{-\frac{1}{2}}\partial_{\mu}\left(\mid g\mid^{\frac{1}{2}}g^{\mu\nu}\partial_{\nu}u\right)-\alpha\frac{\Lambda}{3}u=0,
$$
where $\alpha\in\RR$ is a coefficient linked to the mass ; the equation with
$\alpha=2$ is conformally invariant and corresponds to the massless
case. Using the radial coordinate $x$ given by (\ref{xxx}), we
introduce $f(t,x,\omega):=ru(t\sqrt{\frac{3}{\Lambda}},r,\omega)$ that
is solution of $\partial^2_tf+\mathbf{h}f=0$ with
\begin{equation}
\mathbf{h}:=-\partial_x^2+\frac{2-\alpha}{\cos^2x}-\frac{1}{\sin^2x}\Delta_{S^2_{\omega}}.
  \label{}
\end{equation}
First we investigate the positivity of the potential energy
$$
E(f):=\int_0^{\frac{\pi}{2}}\int_{S^2}
\mid\partial_xf\mid^2+\frac{2-\alpha}{\cos^2x}\mid f\mid^2+\frac{1}{\sin^2x}\mid\nabla_{S^2_{\omega}}f\mid^2dxd\omega.
$$
To estimate the second term, we employ a Hardy inequality. Given
$\phi\in C^1_0([0,\frac{1}{2}[;\RR)$ an integration by part gives
$$
\int_0^{\frac{\pi}{2}}\frac{1}{\cos^2x}\phi^2(x)dx=-\int_0^{\frac{\pi}{2}}2\frac{\phi(x)}{\cos
  x}\phi'(x)\sin xdx\leq\frac{1}{2}\int_0^{\frac{\pi}{2}}\frac{1}{\cos^2x}\phi^2(x)dx+\int_0^{\frac{\pi}{2}}2\phi'^2(x)\sin^2 xdx,
$$
hence
$$
\int_0^{\frac{\pi}{2}}\frac{1}{\cos^2x}\phi^2(x)dx
\leq 4\int_0^{\frac{\pi}{2}}\phi'^2(x)dx.
$$
We deduce that for all $f\in C^{\infty}_0(]0,\frac{\pi}{2}[_x\times S^2_{\omega})$,
$$
<\mathbf{h}f,f>_{L^2}\geq \min(9-4\alpha,1)
\int_0^{\frac{\pi}{2}}\int_{S^2}
\mid\partial_xf\mid^2dxd\omega+
\int_0^{\frac{\pi}{2}}\int_{S^2}
\frac{1}{\sin^2x}\mid\nabla_{S^2_{\omega}}f\mid^2dxd\omega
\geq
 \min(\frac{9}{4}-\alpha,\frac{1}{4})\parallel f\parallel^2_{L^2}
$$
and we conclude that operator $\mathbf h$ endowed with the domain
$D(\mathbf h)= C^{\infty}_0(]0,\frac{\pi}{2}[_x\times S^2_{\omega})$,
is (strictly) positive when $\alpha$
is (strictly) smaller than the upper bound of Breitenlohner-Freedman :
\begin{equation}
\alpha\leq\frac{9}{4}\;\;(respectively\;\;\alpha<\frac{9}{4}).
  \label{}
\end{equation}
We note that for $\alpha=9/4$ and
$f(x,\omega)=\sqrt{\frac{\cos x}{1+\sin x}}$, we have $E(f)=0$,
hence $f\in Ker(\mathbf h^*)\neq\{0\}$.\\

To study the self-adjointness, we expand $f(x,.)$ on the basis of the
spherical harmonics $\left(Y_l^m\right)_{l,m}$ by writting
$$
L^2\left(]0,\frac{\pi}{2}[_x\times
  S^2_{\omega}\right)=\bigoplus_{l=0}^{\infty}\mathbf{L}^2_l,\;\;\mathbf{L}^2_l:=\bigoplus_{m=-l}^{m=l}L^2\left(]0,\frac{\pi}{2}[_x\right)\otimes Y_l^m
$$
therefore $\mathbf h$ is unitarily equivalent to $
\bigoplus_{l=0}^{\infty}\mathbf{h}_l$ where :
$$
\mathbf{h}_l:=-\frac{d^2}{dx^2}+\frac{2-\alpha}{\cos^2x}+\frac{l(l+1)}{\sin^2x},\;\;
D(\mathbf{h}_l)=\bigoplus_{m=-l}^{m=l}C^{\infty}_0(]0,\frac{\pi}{2}[)\otimes Y_l^m.
$$
Since
$\frac{2-\alpha}{\cos^2x}+\frac{l(l+1)}{\sin^2x}-\frac{2-\alpha}{(\frac{\pi}{2}-x)^2}-\frac{l(l+1)}{x^2}$
is a real valued function, bounded on $]0,\frac{\pi}{2}[$, the
symmetric form of the
Kato-Rellich theorem (see \cite{reed}, theorem X.13) assures
that $\mathbf{h}_l$ is essentially self-adjoint iff
$$
\mathbf{k}_l:=-\frac{d^2}{dx^2}+\frac{2-\alpha}{(\frac{\pi}{2}-x)^2}+\frac{l(l+1)}{x^2},\;\;
D(\mathbf{k}_l)=\bigoplus_{m=-l}^{m=l}C^{\infty}_0(]0,\frac{\pi}{2}[)\otimes Y_l^m,
$$
is essentially self-adjoint. By Theorem X.10 of \cite{reed},
$\mathbf{k}_l$ is in the limit point case at zero when $l\geq 1$, and
in the limit point case at $\frac{\pi}{2}$ if $2-\alpha\geq\frac{3}{4}$,
i.e.  $\alpha$ is smaller than the lower bound of Breitenlhoner-Freedman
\begin{equation}
\alpha\leq\frac{5}{4},
  \label{}
\end{equation}
and if $\alpha>\frac{5}{4}$, $\mathbf{k}_l$ is
in the limit circle case at $\frac{\pi}{2}$.
Then the Weyl's limit point-limit circle criterion (see e.g. \cite{pearson},
theorems 6.3 and 6.5), assures that $\mathbf{k}_l$ is essentially self-adjoint when
$l\geq 1$, $\alpha\leq\frac{5}{4}$, and there exists an infinity of
self-adjoint extensions associated with boundary conditions at
$\frac{\pi}{2}$ when $l\geq 1$, $\alpha>\frac{5}{4}$. The case $l=0$ is particular. For
$\alpha<\frac{9}{4}$, the solutions of
$-u''+(2-\alpha)(\frac{\pi}{2}-x)^{-2}u=0$ are
$u=c(\frac{\pi}{2}-x)^{\frac{1}{2}+\sqrt{\frac{9}{4}-\alpha}}+c'(\frac{\pi}{2}-x)^{\frac{1}{2}-\sqrt{\frac{9}{4}-\alpha}}$,
therefore $\mathbf{k}_0$ is
always in the limit circle case at $\frac{\pi}{2}$ and there exists a
lot of self-adjoint extensions. By the
Kato-Rellich theorem (\cite{reed}, theorem X.12), the same results are true for
$\mathbf{h}_l$. Since the spherically symmetric fields play a peculiar
role, we introduce their orthogonal space
$$
\mathbf L^2_{*}:=\left\{f\in L^2(]0,\frac{\pi}{2}[\times S^2);\;\;\forall
  g\in  L^2(]0,\frac{\pi}{2}[),\;\int f(x,\omega)g(x)dxd\omega=0\right\}=\bigoplus_{l=1}^{\infty}\mathbf{L}^2_l,
$$
and $\mathbf h_*$ denotes $\mathbf h$ endowed with the domain
$D(\mathbf h_*)=C^{\infty}_0(]0,\frac{\pi}{2}[\times S^2)\cap \mathbf
L^2_{*}$, and considered as a densely defined operator on $\mathbf L^2_*$. Since this operator is stricly positif when
$\alpha<\frac{9}{4}$, it is essentially selfadjoint iff its range is
dense (\cite{reed}, theorem X.26). We easily prove that 
$\left(Ran(\mathbf{h}_*)\right)^{\perp_{\mathbf
    L^2_*}}=\oplus_{l=1}^{\infty}\left(Ran(\mathbf{h}_l)\right)^{\perp_{\mathbf L^2_l}}$, and we conclude that $\mathbf h_*$ is essentially self-adjoint when $\alpha\leq\frac{5}{4}$. Finally we have proved the following :
\begin{Theorem} 
When $\alpha\leq\frac{9}{4}$ (respect. $\alpha<\frac{9}{4}$),
$\mathbf{h}$ is a  positive (respect. strictly positive) symmetric operator on
$L^2(]0,\frac{\pi}{2}[\times S^2)$. When
$\frac{5}{4}<\alpha<\frac{9}{4}$ there exists an infinity of
self-adjoint extensions of $\mathbf{h}_*$ on $\mathbf L^2_*$, associated with boundary
conditions on $\{\frac{\pi}{2}\}\times S^2$. When
$\alpha\leq\frac{5}{4}$,  $\mathbf{h}_*$ is essentially self-adjoint.
  \label{}
\end{Theorem}
%%%%%%%%%%%%%%%%%%%%%%%%%%%%%%%%%%%%%%%%%%%%%%%%%%%%%%%%%%%%%%%%%%%%%%%%%%%%%%%%%%%%
%%%%%%%%%%%%%%  BIBLIOGRAPHY   %%%%%%%%%%%%%%%%%%%%%%%%%%%%%%%%%%%%%%%%%%%%%%

%%%%%%%%%%%%%%%%%%%%%%%%%%%%%%%%%%%%%%%%%%%%%%%%%%%%%%%%%%%%%%%%%%%%


\begin{thebibliography}{10}

\bibitem{avis}
S.J.~{\sc{Avis, C.J. Isham, D. Storey}}.
\newblock \protect{Quantum field theory in anti-de Sitter space-time}.
\newblock {\em Phys. Rev. D},
 18 (10) : 3565--3576, 1978.

\bibitem{ba1}
A.~{\sc{Bachelot}}.
\newblock \protect{Global properties of the wave equation on non globally hyperbolic manifolds}.
\newblock {\em J. Math. Pures Appl.},
 81 : 35--65, 2002.

\bibitem{equipart}
A.~{\sc{Bachelot}}.
\newblock \protect{Equipartition de l'\'energie pour les syst\`emes
  hyperboliques et formes compatibles}.
\newblock {\em Ann. Inst. Henri Poincar\'e - Physique th\'eorique},
  46(1) : 45--76, 1987.

\bibitem{bachelot-motet}
A. {\sc{Bachelot-Motet}}.
\newblock \protect {Nonlinear Dirac fields on the Schwarzschild metric}.
\newblock {\em Class. Quantum Grav.}, 15, 1815--1825, 1998.

\bibitem{bartnik}
R. A.~{\sc Bartnik, P. T. Chru\'sciel}.
\newblock \protect{Boundary value problems for Dirac equations with applications}.
\newblock{\em  
J. Reine Angew. Math.}, 579, 13--73,
2005.

\bibitem{booss}
B.~{\sc Boo}{\ss}, {\sc K. P. Wojciechowski}.
\newblock \protect{Elliptic boundary problems for Dirac Operators}.
\newblock Birkh\"auser, 1993.

\bibitem{breit1}
P.~{\sc Breitenlohner, D. Z. Freedman}.
\newblock \protect{Stability in gauged extended supergravity}.
\newblock{\em Ann. Phys.},
144, 2, 249--281, 1982.

\bibitem{breit2}
P.~{\sc Breitenlohner, D. Z. Freedman}.
\newblock \protect{Positive energy in anti-de Sitter backgrounds and gauged extended supergravity.}
\newblock{\em Phys. Lett. B},
115, 3,  197--201, 1982.

\bibitem{bruning}
J.~{\sc Br\"uning, M. Lesch}.
\newblock \protect{On Boundary Value Problems for Dirac Type Operators
I. Regularity and Self-Adjointness}.
\newblock{\em J. Func. Anal.},
185, 1--62, 2001.

\bibitem{bunke}
U.~{\sc Bunke}.
\newblock \protect{Comparison of Dirac operators on manifolds with boundary}.
\newblock{\em Suppl. di Rend. Circ. Mat. Palermo, Serie II},
133--141, 1993.

\bibitem{choquet}
Y. {\sc{Choquet-Bruhat}}.
\newblock \protect {Solutions globales d'\'equations d'ondes sur
  l'espace-temps Anti de Sitter}.
\newblock {\em C. R. Acad. Sci. Paris}, 308, 323--327, 1989.

\bibitem{cotaescu}
I. I. {\sc{Cot\u{a}escu}}.
\newblock \protect {Normalized energy eigenspinors of the Dirac field on anti-de Sitter spacetime}.
\newblock {\em Phys. Rev. D}(3) 60, no. 12, 124006, 4pp, 1999.

\bibitem{fr2}
J. L. ~{\sc{Friedman, M.S. Morris}}.
\newblock \protect{Existence and Uniqueness Theorems for Massless Fields on a Class of Spacetimes with Closed Timelike Curves}.
\newblock {\em Commun. Math. Phys.},
  186: 495--529, 1997.


\bibitem{gel}
I. M. {\sc{Gelfand, R. A. Minlos, Z. Ya. Shapiro}}.
\newblock \protect{Representations of the rotation and Lorentz groups
  and their representations}.
\newblock Pergamon Press, 1963.


\bibitem{gibbons}
G.W.~{\sc Gibbons}.
\newblock \protect{Anti-de-Sitter spacetime and its uses},
in
\newblock{\it Mathematical and quantum aspects of relativity and
  cosmology (Pythagoreon, 1998)},
\newblock{\em Lecture Notes in Phys.}, 537, Springer-Verlag, 102--142, 2000.



\bibitem{grubb}
G.~{\sc Grubb}.
\newblock \protect{Spectral Boundary Conditions for Generalizations of
  Laplace and Dirac Operator}.
\newblock{\em Comm. Math. Phys.},
240,  243--280, 2003.

\bibitem{haf}
D. {\sc{H\"afner}}.
\newblock \protect{Creation of fermions by rotating charged black-holes}.
arXiv:math/0612501v1,
115 pp., 2006.

\bibitem{ha-n}
D. {\sc{H\"afner}, J-P. Nicolas}.
\newblock \protect{Scattering of massless Dirac fields by a Kerr black hole}.
\newblock {\em Reviews in Math. Physics},
29--123, 2007.

\bibitem{haw}
S. W. {\sc{Hawking, G. F. R. Ellis}}.
\newblock \protect{The large scale structure of space-time}.
\newblock Cambridge University Press, 1973.

\bibitem{hijazi}
O.~{\sc Hijazi, S. Montiel, A. Roldan}.
\newblock \protect{Eigenvalue Boundary Problems for the Dirac Operator}.
\newblock{\em Comm. Math. Phys.},
231,  375--390, 2002.

\bibitem{ishi1}
A. {\sc{Ishibashi, R. M. Wald}}.
\newblock \protect {Dynamics in non-globally-hyperbolic, static
    space-times : II. General analysis of prescriptions for dynamics}.
\newblock {\em Class. Quantum Grav.}, 20, 3815--3826, 2003.

\bibitem{ishi2}
A. {\sc{Ishibashi, R. M. Wald}}.
\newblock \protect {Dynamics in non-globally-hyperbolic, static
    space-times : III. Anti-de-Sitter space-time}.
\newblock {\em Class. Quantum Grav.}, 21, 2981--3013, 2004.


\bibitem{kalf}
H.  {\sc{Kalf, O. Yamada}}.
\newblock \protect{Essential self-adjointness of Dirac operators with
  a variable
  mass}.
\newblock {\em Proc. Japan Acad. Ser. A}, 
76, 2, 13--15, 2000.

\bibitem{lions}
J-L.  ~{\sc{Lions, E. Magenes}}.
\newblock {\em Probl\`emes aux limites non homog\`enes et applications I}.
\newblock {Dunod}, 1968.

\bibitem{melnyk}
F.~{\sc Melnyk}.
\newblock \protect{Scattering on Reissner-Nordstr{\o}m metric for massive charged spin 1/2 fields}.
\newblock{\em Ann. Henri Poincar\'e},
4,  $n^o$5,  813--846, 2003.


\bibitem{JPN-Dirac}
J-P. {\sc{Nicolas}}.
\newblock \protect{Scattering of linear Dirac fields by a spherically symetric
  Black-Hole}.
\newblock {\em Ann. Inst. Henri Poincar\'e - Physique th\'eorique},
  62,  $n^o$2, 145--179, 1995.

\bibitem{JPNdiss}
J-P. {\sc{Nicolas}}.
\newblock \protect{ Dirac fields on asymptotically flat space-times}.
\newblock {\em Dissertationes Math.},
  408, 2002.

\bibitem{JPN3/2}
J-P. {\sc{Nicolas}}.
\newblock \protect{Global exterior Cauchy problem for the spin 3/2
  zero rest-mass fields in the Schwarzschild space-time}.
\newblock {\em Comm. Partial Differential Equations},
  22,  $n^o$3-4, 465--502, 1997.

\bibitem{oneill}
B. {\sc{O'Neill}}.
\newblock \protect{Semi-Riemannian geometry. With Applications to Relativity}.
\newblock {\em  Pure and Applied Mathematics, 103},
Academic Press, 1983.

\bibitem{pearson}
D. B.  ~{\sc{Pearson}}.
\newblock {\em Quantum scattering and spectral theory}.
\newblock {Academic Press}, 1988.

\bibitem{reed}
M. {\sc{Reed, B. Simon}}.
\newblock \protect{Methods of modern mathematical physics,
vol. 2, Fourier Analysis, Self-Adjointness}.
\newblock
Academic Press, 1975.


\bibitem{segev}
I. {\sc{Segev}}.
\newblock \protect {Dynamics in stationary, non-globally hyperbolic spacetimes}.
\newblock {\em Class. Quantum Grav.}, 21, 2651--2668, 1994.

\bibitem{shishkin}
G. V.  {\sc{Shishkin, V. M. Villalba}}.
\newblock \protect{Dirac Equation in external vector fields :
  separation of variables}.
\newblock {\em J. Math. Phys.}, 
30, 2132--2143  , 1989.

\bibitem{schmidt}
K. M.  {\sc{Schmidt, O. Yamada}}.
\newblock \protect{Spherically symmetric Dirac operators with variable
  mass and potential infinite at infinity}.
\newblock {\em Publ. Res. Inst. Math. Sci. Kyoto Univ.}, 
34, 211--227, 1998.

\bibitem{vil}
N. J. {\sc{Vilenkin}}.
\newblock \protect{Special Functions and the Theory of Group Representations}.
\newblock {\em Translations of Mathematical Monographs, volume 22, A.M.S.},
1968.

\bibitem{vilk}
N. J. {\sc{Vilenkin, A. U. Klimyk}}.
\newblock \protect{Representation of Lie Groups and Special Functions,
vol. 1}.
\newblock Mathematics and Its Applications (Soviet Series), Kluwer
Academic Publishers, 1991.

\bibitem{wald}
R. M.  {\sc{Wald}}.
\newblock \protect{Dynamics in nonglobally hyperbolic, static space-times}.
\newblock {\em J. Math. Phys.}, 
21 (12), 2802--2805, 1980.

\bibitem{yamada}
O.  {\sc{Yamada}}.
\newblock \protect{On the spectrum of Dirac operators with the
  unbounded potential at infinity}.
\newblock {\em Hokkaido Math. J.}, 
26, 439--449, 1997.



\end{thebibliography}
\end{document}